\documentclass[fleqn,usenatbib,useAMS]{mnras}

\usepackage{graphicx}% Including figure files
\graphicspath{ {LEGACY/images/} }
\usepackage{caption}
\usepackage{subcaption}
\usepackage{amsmath}	% Advanced maths commands
\usepackage{amssymb}	% Extra maths symbols
\usepackage{multicol}        % Multi-column entries in tables
\usepackage{bm}		% Bold maths symbols, including upright Greek
\usepackage{pdflscape}	% Landscape pages
\usepackage{rotating}
\usepackage{longtable}
\usepackage[utf8]{inputenc}
\usepackage{threeparttable}
\usepackage[T1]{fontenc}
\usepackage{ae,aecompl}

\usepackage{newtxtext,newtxmath}
\usepackage{lineno}

\title[A Multi-wavelength search for Black Widows and Redbacks]{\textit{A Multi-wavelength search for Black-Widow and Redback counterparts of candidate $\gamma$-ray millisecond pulsars}}

\author[C. Braglia et al.]{C. Braglia$^{1}$,
R. P. Mignani$^{2,3}$\thanks{Contact e-mail:\href{mailto:roberto.mignani@inaf.it}{roberto.mignani@inaf.it}},
A. Belfiore$^{2}$,
M. Marelli$^{2}$,
G. L. Israel$^{4}$,
G. Novara$^{2,5}$,
\newauthor{
A. De Luca$^{2,6}$,
A. Tiengo$^{2,5,6}$,
P. M. Saz Parkinson$^{7,8}$
}
\\
\\
$^{1}$ University of Milan, Department of Physics "A. Pontremoli", via G. Celoria 16, 20133, Milan, Italy \\
$^{2}$ INAF - Istituto di Astrofisica Spaziale e Fisica Cosmica Milano, via A. Corti 12, 20133, Milan, Italy\\
$^{3}$ Janusz Gil Institute of Astronomy, University of Zielona G\'ora, ul Szafrana 2, 65-265, Zielona G\'ora, Poland \\
$^{4}$ INAF–Osservatorio Astronomico di Roma, via Frascati 33, I-00040 Monteporzio Catone, Italy \\
$^{5}$ Scuola Universitaria Superiore IUSS Pavia, Piazza della Vittoria 15, I-27100 Pavia, Italy \\
$^{6}$ INFN, Sezione di Pavia, via A. Bassi 6, I-27100 Pavia, Italy\\
$^{7}$ Santa Cruz Institute for Particle Physics, University of California, Santa Cruz, CA 95064, USA \\
$^{8}$ Department of Physics, Laboratory for Space Research, University of Hong Kong, Pokfulam Road, Hong Kong 
 }
 
\date{Last updated 2015 May 22; in original form 2013 September 5}

\pubyear{2015}

\begin{document}
\label{firstpage}
\pagerange{\pageref{firstpage}--\pageref{lastpage}}
\maketitle

\begin{abstract}
The wealth of detections of millisecond pulsars (MSPs) in $\gamma$-rays by {\em Fermi} has spurred searches for these objects among the several unidentified $\gamma$-ray sources. Interesting targets are a sub-class of binary MSPs, dubbed "Black Widows" (BWs) and "Redbacks" (RBs), which are in orbit with low-mass non-degenerate companions fully or partially ablated by irradiation from the MSP wind. These systems can be easily missed in radio pulsar surveys owing to the eclipse of the radio signal by the intra-binary plasma from the ablated companion star photosphere, making them better targets for multi-wavelength observations. We used optical and X-ray data from public databases to carry out a systematic investigation of all the unidentified $\gamma$-ray sources from the Fermi Large Area Telescope (LAT) Third Source Catalog (3FGL), which have been pre-selected as likely  MSP candidates according to a machine-learning technique analysis. We tested our procedure by recovering known binary BW/RB identifications and searched for new ones, finding two possible candidates. At the same time, we investigated previously proposed BW/RB identifications and we ruled out one of them based upon the updated $\gamma$-ray source coordinates.
\end{abstract}


%%%%%%%%%%%%%%%%%%%%%%%%%%%%%%%%%%%%%%%%%%%%%%%%%%

%%%%%%%%%%%%%%%%% BODY OF PAPER %%%%%%%%%%%%%%%%%%

% The MNRAS class isn't designed to include a table of contents, but for this document one is useful.
% I therefore have to do some kludging to make it work without masses of blank space.
%\begingroup
%\let\clearpage\relax
%\tableofcontents
%\endgroup
%\newpage

\section{Introduction}
\label{subsec:intro}

Millisecond pulsars (MSPs) are a sub-group of radio pulsars characterised by much shorter spin
periods ($P_{\rm s}\la$10 ms) than the rest of the population. They are also very stable clocks, with spin period derivatives $\dot{P_{\rm s}} \sim 10^{-21}$--$10^{-18}$s s$^{-1}$, which, according to the magnetic dipole model, imply characteristic ages $P_{\rm
s}/2\dot{P_{\rm s}} \sim$1--10 Gyrs, low dipolar magnetic fields of $B \sim 10^{8}$--$10^{9}$ G, and rotational energy loss rates as low as $\dot{E} \sim 10^{32}$--$10^{36}$ erg s$^{-1}$.
According to the canonical recycling scenario \citep{Alpar1982, Radha1982}, the short spin periods of MSPs are explained by a phase of matter accretion from a non-degenerate companion star, with the consequent spin up of the neutron star. This scenario is rooted in the fact that the majority of MSPs (214/333\footnote{{\tt https://www.atnf.csiro.au/research/pulsar/psrcat/} - v1.63}) are indeed observed in binary systems, usually with a white dwarf (WD) companion peeled off of its external layers. 

The existence of {\em isolated} MSPs in the Galactic field \citep[e.g., PSR\, B1937+21;][]{Backer1982}
then represented for a long time a conundrum. This was solved by the discovery of the MS PSR\, B1957+20 \citep{Fruchter1988} in orbit around a very low-mass ($\sim 0.002 M_{\odot}$) companion, with the radio signal regularly disappearing during part of the orbit. This behaviour was interpreted as an eclipse of the radio signal absorbed or scattered by intra-binary gas produced from ablation of the companion star irradiated by the pulsar wind, hence leading to the nickname  "Black Widow" (BW) for this pulsar. Isolated MSPs were then naturally explained as descendants of families of BW pulsars, where the companion has been fully ablated, eventually. In addition to BWs, other similar systems were discovered, starting with PSR\, J1023+0038 \citep{Archibald2009}, where the companion star has masses $M_{\rm C}\sim$0.1--0.4$M_{\odot}$ and is only partially ablated by irradiation from the pulsar wind.  Following the spider analogy, these systems were nicknamed "Redbacks" \citep[RBs;][]{Roberts2012}, after the species of Australian spiders where females feed only part of their lighter male companions after mating.
Both systems are characterised by very short orbital periods (P$_{\rm B}<$1 d).

 These "spiders" are crucial to understand the MSP recycling scenario and the formation of isolated MSPs, study the acceleration, composition and shock dynamics of the MSP winds, and infer accurate MSP mass measurements through pulse timing, key to determine the neutron star equation of state \citep[see, e.g.][for a recent review]{Linares:2019aua}.
 They are elusive targets in radio pulsar surveys owing to the radio signal eclipse, with the eclipse extent unknown and variable in time. Furthermore, some  RBs alternate between accretion-powered states, with high X-ray emission and no radio pulsations, and rotation-powered states, with low X-ray emission and radio pulsations \citep{Linares2014}.  The study of these "transitional" MSPs, with only three such systems known in the Galactic field \citep[e.g.,][for a recent review] {Jaodand2018}, is key to track the long-sought evolutionary link between accreting neutron stars in Low Mass X-ray Binaries and MSPs in binary systems.  
 
 Since MSPs (regardless of the type) are almost half of the $\gamma$-ray pulsar population (115 out of 250, of which 91 in binary systems\footnote{{\tt https://confluence.slac.stanford.edu/display/GLAMCOG/\\Public+List+of+LAT-Detected+Gamma-Ray+Pulsars}}), many of the "spiders" known to date in the Galactic field, 43 confirmed as radio/$\gamma$-ray pulsars and 11 candidates  \citep{Linares:2019aua}, have been searched for in unidentified $\gamma$-ray sources discovered by the {\em Fermi} Large Area Telescope \citep[LAT;][]{Atwood2009}. Candidate "spiders" are usually pinpointed through multi-wavelength follow-up observations which trigger dedicated radio pulsar searches, with $\gamma$-ray pulsations searched for using the radio pulsar ephemerides, like in the case of PSR\, J2339$-$0533 \citep{Ray2020}. Much more rarely, the detection of $\gamma$-ray pulsations \citep{Pletsch2012, Clark2017} directly triggers radio follow-ups \citep{Ray2013}. As of now, 37 "spiders" have been detected and seen to pulsate in $\gamma$-rays \citep{Hui2019}. In many cases, optical  observations have been key to finding  BW/RB candidates via the discovery of $\la1$d periodic flux modulations from the tidally distorted and irradiated MSP companion, which traces the binary system orbital period \citep[see][and references therein]{Salvetti2015}  and facilitates radio/$\gamma$-ray pulsation searches. 
 The quest for "spiders" among unidentified {\em Fermi} sources  is still restlessly pursued, with promising candidates for multi-wavelength observations selected from the similarity between the $\gamma$-ray source temporal and spectral characteristics to those of known $\gamma$-ray MSPs.  
 
 In this work\footnote{A more extended description can be found in C. Braglia (2020), MSc Thesis Dissertation, University of Milan, Italy}, we assumed as a reference the candidate MSP selection done by \citet{SazParkinson2016} from a machine-learning analysis of unidentified $\gamma$-ray sources in the Fermi Large Area Telescope (LAT) Third Source Catalog \citep[3FGL;][]{Acero2015}. For consistency, at this stage we did not include MSP candidates selected by independent statistical analysis of unidentified 3FGL sources, e.g. \citet{Dai2016, Dai2017}.
 %; \citet{Kaur:2019cql}). 
 %
 Our strategy is described in Sectn.\, \ref{subsec:strat}, whereas the multi-wavelength analysis is described in Sectn.\, \ref{subsec:corr} with the results presented and  discussed in Sectn.\, \ref{subsec:results}. Summary and conclusions follow in Sectn.\ \ref{subsec:summ}.
 
\section{Strategy}
\label{subsec:strat}

\subsection{Candidate Selection}
The pulsar-like candidate list of \citet{SazParkinson2016} includes 120 unidentified 3FGL $\gamma$-ray sources classified in two groups: young pulsars (YNG) and MSPs.  To be conservative, we considered candidates with either a non-ambiguous ("MSP/MSP"; 41) or a tentative ("MSP/YNG" or "YNG/MSP"; 7)  MSP classification. Of course, we cannot rule out that apparent mis-classifications between the two groups lead to the loss of genuine MSP candidates and to the acquisition of false ones.  However, by training their algorithms on a sample of identified/associated $\gamma$-ray sources, \citet{SazParkinson2016} claim an overall classification accuracy of $\ga$96\%, so that we expect the number of apparent mis-classifications to be limited to very few cases only. We note that since the machine learning techniques used by \citet{SazParkinson2016} cannot distinguish between isolated and binary MSPs on the basis of their $\gamma$-ray temporal and spectral characteristics alone,  our starting sample may very well include both isolated and binary MSP candidates. Moreover, no periodic $\gamma$-ray modulations associated with the orbital period of a compact binary system, like e.g. in 3FGL\, J2039.6$-$5618 \citep{Ng2018}, or $\gamma$-ray state transitions, like in some RBs \citep{Torres2017}, have been recognised yet for the vast majority of the above 48 MSP candidates. Therefore, we are left with no option other than applying our identification strategy, tailored on BWs and RBs, to all the 48 MSP candidates.

To test and validate our strategy we did not filter out from our starting sample those $\gamma$-ray sources already proposed as strong BW/RB candidates prior to the publication of the work of \citet{SazParkinson2016}, such as 
3FGL\, 2039.6$-$5618 \citep{Salvetti2015}, 
or proposed afterwards, such as 
3FGL\, 0954.8$-$3948 \citep{Li2018}, 
and those unambiguously identified as pulsars by the discovery of radio and/or $\gamma$-ray pulsations, such as 3FGL\, 1946.4$-$5403 \citep{Camilo2015,Ray2016}. For the same reason, we did not filter out sources which have been eventually identified as young pulsars and not as MSPs, such as 4FGL\, J0359.4+5414 \citep{Clark2017}, which had a tentative "MSP/YNG" classification in \citet{SazParkinson2016}, and we used them as false positives.
In order to avoid missing potentially interesting candidates, we did not apply a cut in the MSP candidate classification ranking and we did not filter them according to their 95\% confidence error radius (r95). Although MSPs are expected to be found mostly at high Galactic latitudes, where they migrate owing to their proper motion in their Gyr-long lifetimes, their orbital motion in the Galactic potential can bring them back to the Galactic plane. Therefore, we did not apply a sample selection based on the source coordinates.

\subsection{Multi-wavelength Database}
The multi-wavelength identification strategy is the same as described in \citet{Salvetti2017} and applied to other BW/RB searches. Briefly, since BWs/RBs are also observed in the optical/X-rays \citep[see, e.g.][for a compilation]{Hui2019}, this strategy consists of i) mapping the $\gamma$-ray source error box in X-rays to spot candidate counterparts, ii) looking for optical counterparts to the selected X-ray sources, iii) searching for optical modulations with periods $\la$1 d. 

Since our starting sample includes 48 $\gamma$-ray sources, with a random distribution in right ascension and declination, running the multi-wavelength identification effort through dedicated follow-up X-ray/optical observations for them all would be unrealistic in terms of required observing time and data analysis time investment. Therefore, we used data collected in public archives and the derived source catalogues and data products, as done in \citet{Mignani2014}. In particular, in the X-rays we used the third {\em XMM-Newton} serendipitous source catalogue \citep{Rosen2016} Data Release 8 (3XMM-DR8), the {\em Chandra} Source Catalogue \citep{Evans2010} Release 2.0 (CSC\, 2.0), and the {\em Swift} X-ray Telescope (XRT) Point Source Catalogue
\citep[1SXPS;][] {Evans2013}. Owing to the limited number of sources (497) and to the extragalactic nature of about half of them, we did not use the {\em NuSTAR} Serendipitous Survey 40-month catalog \citep{Lansbury2017}. Since we aim at looking for orbital modulations from the optical counterparts to the X-ray sources, we used multi-epoch imaging data and associated catalogues from wide-area optical sky surveys. These are the Catalina Sky Survey (CSS), assembling under the same name the original CSS \citep{Larson2003} plus its siblings the Mount Lemmon Survey (MLS) and the Siding Spring Survey (SSS), the Palomar Transient Factory \citep[PTF;][]{Rau2009}, the intermediate Palomar Transient Factory \citep[iPTF;][]{Kulkarni2013}, the Zwicky Transient Factory \citep[ZTF;][]{Bellm2018} surveys, and the Pan-Starss survey \citep{Chambers2016}.  In all cases, we used the most recent survey and data products releases. We decided not to use the recently released  Hubble Catalog of Variables \citep{Bonanos2019} because of the small sky coverage and non-uniform cadence of the multi-epoch {\em HST} images.

Both in the X-rays and in the optical, the choice of different catalogues/surveys is dictated by their complementarity. For instance, observations with {\em Swift} provide exposures for several unidentified $\gamma$-ray source fields, although they are relatively shallow, whereas observations with {\em XMM-Newton} provide a much sparser $\gamma$-ray source field mapping but are much deeper. Owing to the smaller field of view of its detectors, observations with {\em Chandra} have been rarely used to map $\gamma$-ray source fields but, when used, provide a much more accurate source positioning than both {\em Swift} and {\em XMM-Newton}, crucial for the X-ray source optical counterpart identification. 
In the optical, the use of different surveys yields complementary sky mappings, extended multi-band flux information, more sensitive flux limits, and finer time resolution related to the different observing cadence. In addition, for sources with an already known orbital period, the use of different surveys could help detecting period variations over time, which are expected for BWs and RBs owing to the mass loss from the companion star caused by the MSP irradiation, by comparing light curves obtained at different epochs \citep[e.g.,][]{Cho2018}. Furthermore, different surveys also allow us to study variations in the light curve profile as observed in different optical bands,  such as in PanSTARRS,  which are expected as the result of the companion star irradiation.  

Of course, owing to the non-homogeneous sky mapping possible with the available survey data (especially in the X-rays, given the serendipitous nature of the surveys) some of the 48 $\gamma$-ray source fields might suffer of uneven or no multi-wavelength coverage. Moreover, the catalogue releases are based on observations processed up to a certain date. This means that multi-wavelength data products for a given field might not be available yet. Owing to the project's time constraints, we did not process optical/X-ray data not yet included in the most recent releases of the reference catalogues and we did not scan the entirety of public optical data archives for unprocessed data. We considered such options only on a case by case basis, depending on the preliminary results of our analysis.

Our work sets the state of the art of the multi-wavelength investigation of all candidate MSPs selected from the 3FGL by \citet{SazParkinson2016}, here carried out systematically for the first time.  Some of these candidates have been already investigated by \citet{Salvetti2017} and here we have complemented their analysis using an extended multi-wavelength database. 

\section{Multi-wavelength analysis}
\label{subsec:corr}

\subsection{Coordinate re-assessment}
\label{subsec:coo}

Before searching for X-ray and optical counterparts, we first obtained the $\gamma$-ray source positions using the recently released Fermi Large Area Telescope Fourth Source Catalog \citep[4FGL;][]{4fgl} based on the first eight years of data acquisition of the {\em Fermi} mission, which supersedes the Fermi LAT 8-Year Source Catalog published in 2018. We used the latest 4FGL version, released on 15 May 2019. 
The 4FGL data re-processing with the {\tt Pass 8} software \citep{Bruel2018} allowed to obtain a more precise $\gamma$-ray coordinate determination and a smaller $\gamma$-ray error ellipse, which could lead to the discovery of new X-ray/optical counterparts to the $\gamma$-ray source, previously incompatible with the actual $\gamma$-ray source position. A clear example can be found in the work of \citet{Li2018}, where the X-ray and optical counterparts to 3FGL\, J0954.8$-$3948 were only found after the relocalization of the $\gamma$-ray source, offset by $\sim 7\arcmin$ with respect to its original 3FGL coordinates.

\begin{figure*}
\includegraphics[width=0.8\textwidth]{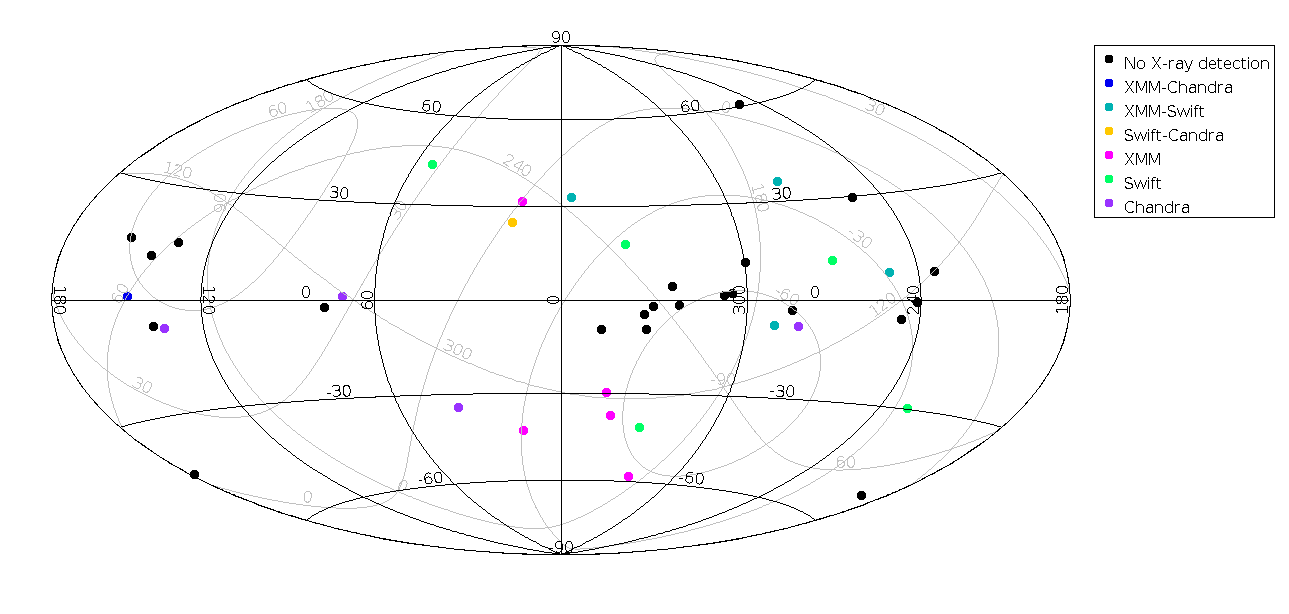}
\caption{Sky map in Galactic coordinates (black solid lines) with the positions of the 48 $\gamma$-ray MSP candidates of \citet{SazParkinson2016} marked with different colors depending on the X-ray coverage from different surveys (see legenda). $\gamma$-ray sources with no X-ray coverage in any of the reference surveys are plotted in black. As it can be seen, these sources are mostly distributed along the Galactic plane, which may reflect an observational bias in the coverage of these crowded X-ray source regions. }
\label{fig:skymap}
\end{figure*}

Since from now on we use the 4FGL coordinates as a reference for our work, we checked  how many of the MSP candidates of \citet{SazParkinson2016}, selected from the 3FGL, are still flagged unassociated in the 4FGL. We found that nine of them have a pulsar association in the 4FGL. Two of them, 4FGL\, J1625.1$-$0020 \citep{Salvetti2017} and 4FGL\, J1653.6$-$0158 \citep{Romani2014}, however,  should still be regarded as candidates since the discovery of pulsations has not been reported yet in a dedicated paper. 
Of the remaining seven which are confirmed pulsars,
4FGL\, J0318.2+0254 \citep[PSR\, J0318+0253;][]{Wang2018}, 4FGL\, J0359.4+5414 \citep[PSR\, J0359+5414;][]{Clark2017}, 4FGL\, J1035.4$-$6720 \citep[PSR\, J1035$-$6720;][]{Clark2018}, 4FGL\, J1528.4$-$5838 \citep[PSR\, J1528$-$5838;][]{Clark2017}, 4FGL\, J1641.2-5317 \citep[PSR\, J1641$-$5317;][]{Clark2017}, 4FGL\, J1744.0$-$7618 \citep[PSR\, J1744$-$76194][]{Clark2018}, and 4FGL\, J1946.5-5402 (PSR\, J1946$-$5403),
 only the last one  is a binary MSP \citep{Camilo2015}. However, like we stated in Sectn.\, \ref{subsec:strat}, we kept these sources in our sample as a blind test of our procedure.

\subsection{Cross-correlation with X-ray catalogues}
\label{subsec:xcorr}

First of all, we searched for candidate X-ray counterparts to the 48 $\gamma$-ray sources performing a cross-match with {\em XMM-Newton}, {\em Chandra} and {\em Swift} sources that lie inside the 95$\%$ confidence 4FGL error ellipse. 
We performed the cross-match considering only sources that are present in the most recent releases of the X-ray catalogues publicly available. Like we explained in Sectn.\, \ref{subsec:strat}, since each of them is based on a number of years of observations, 
we did not include sources which have been detected in observations performed after the time span covered by the catalogue.

We found that among the 48 $\gamma$-ray sources only  23 have at least an X-ray counterpart candidate inside the 95\%-confidence $\gamma$-ray source error ellipse from at least one of the three X-ray surveys (Fig.\,\ref{fig:skymap}). 
The results of the cross-matches with the X-ray catalogues for these 23 $\gamma$-ray sources are visualised in more detail in Fig.\,\ref{fig:coverage},\ref{fig:coverage2} and summarised in columns 3--5 of Table\, \ref{tab:my-table}.
The majority of X-ray candidate counterparts (41) comes from the 3XMM/DR8 catalogue, while 14 come from the 1SXPS and 19 from the CSC\, 2.0. The relatively small number of X-ray sources detected by each facility is explained both by the random coverage of each $\gamma$-ray error ellipse and the random distribution of the X-ray integration times (Fig.\, \ref{fig:histocover}).

\renewcommand{\thefigure}{2a}
\begin{figure*}
\centering
\begin{tabular}{ccc}
{\includegraphics[width=0.32\textwidth]{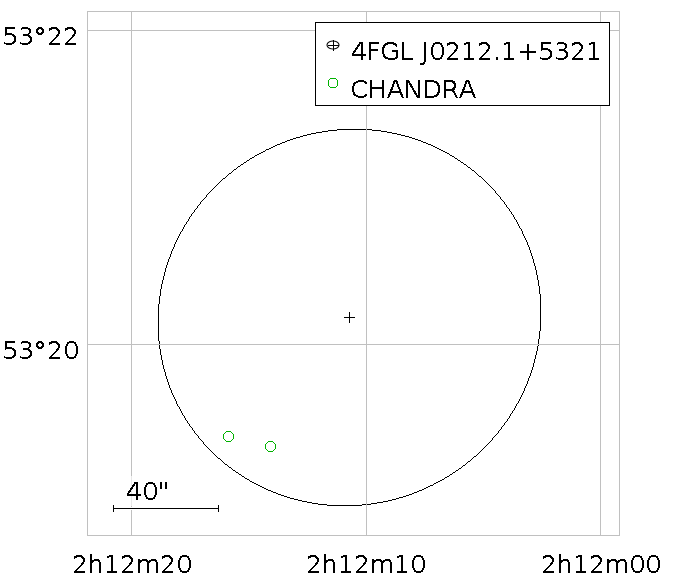}} &
{\includegraphics[width=0.32\textwidth]{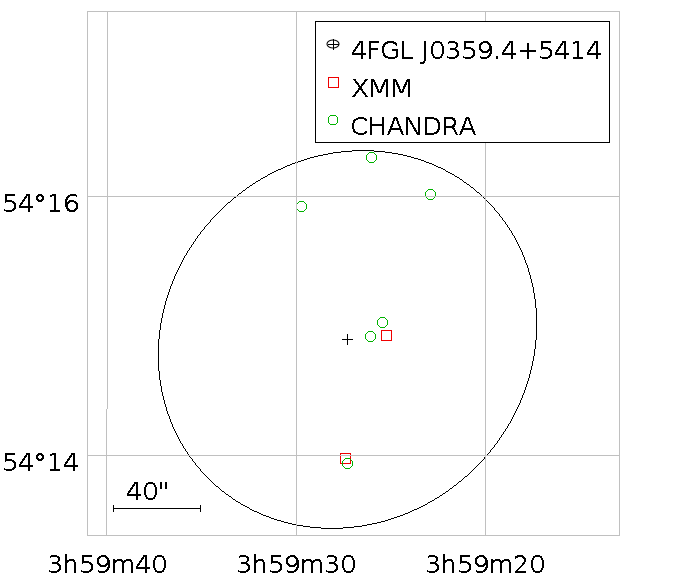}} &
{\includegraphics[width=0.32\textwidth]{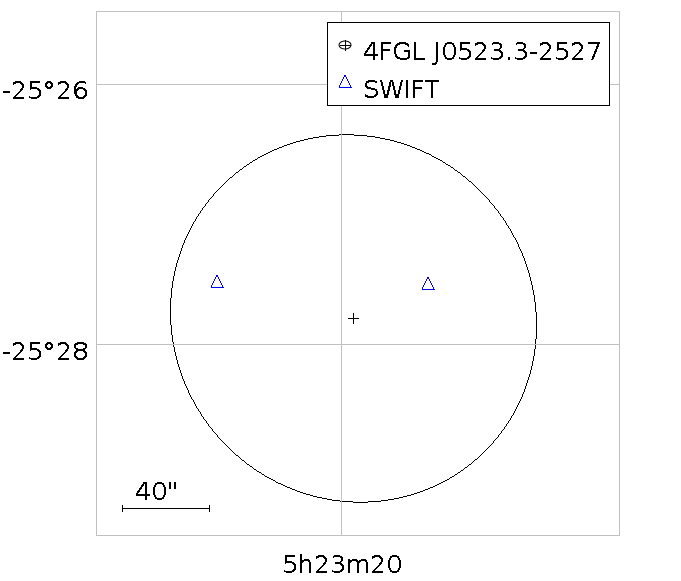}} \\
{\includegraphics[width=0.32\textwidth]{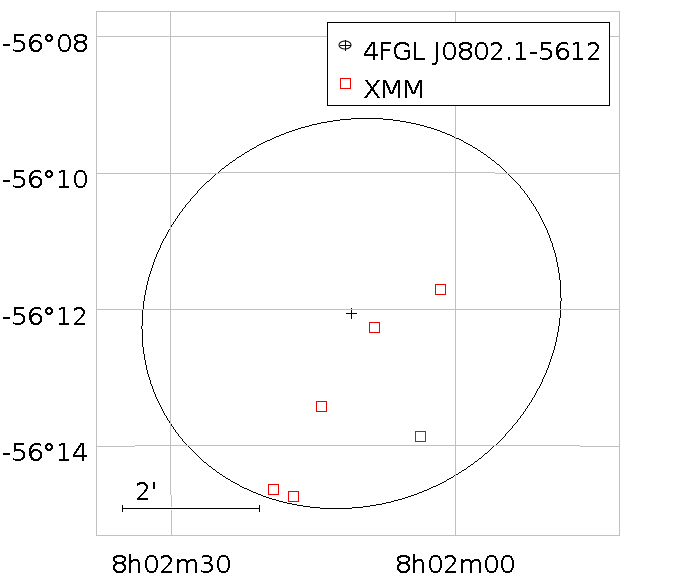}} &
{\includegraphics[width=0.32\textwidth]{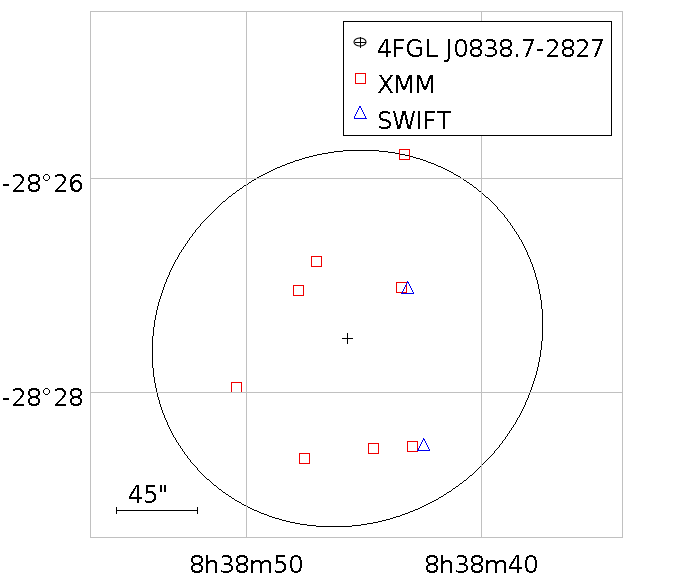}} &
{\includegraphics[width=0.32\textwidth]{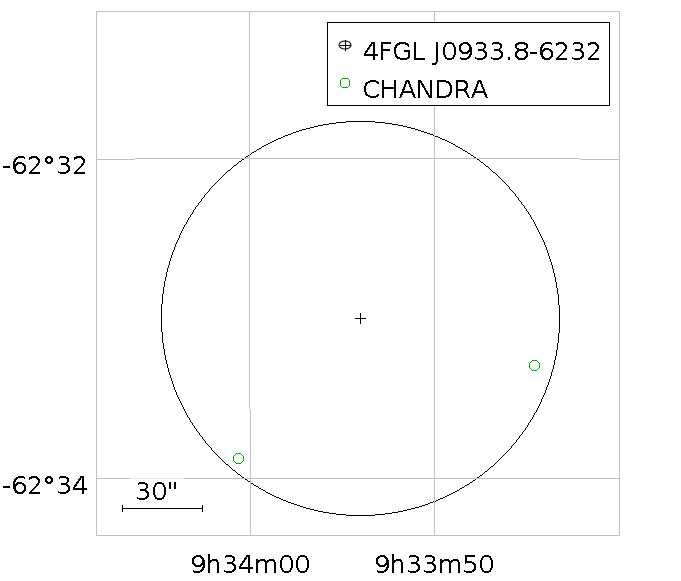}} \\
{\includegraphics[width=0.32\textwidth]{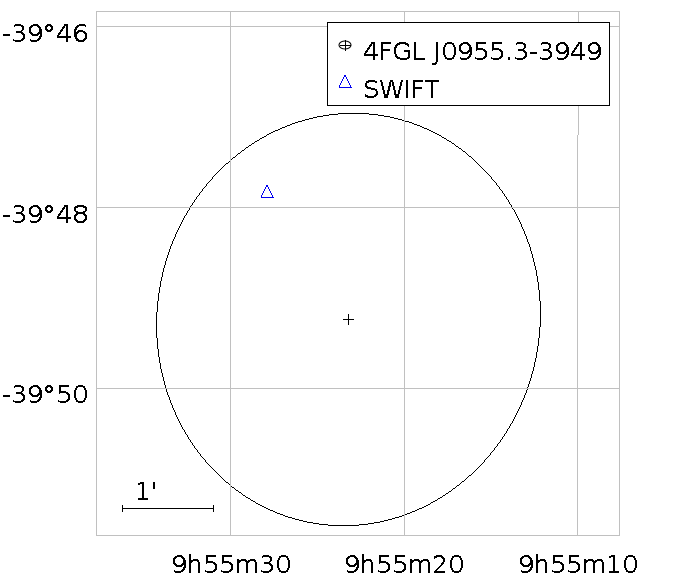}} &
{\includegraphics[width=0.32\textwidth]{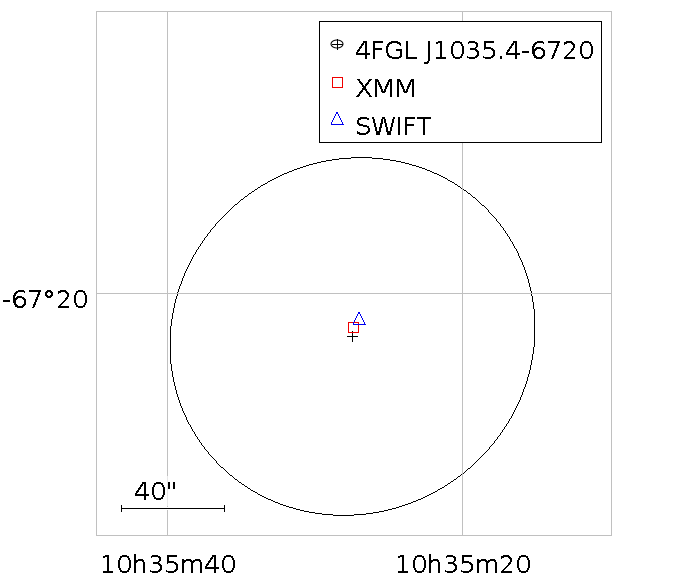}}&
{\includegraphics[width=0.32\textwidth]{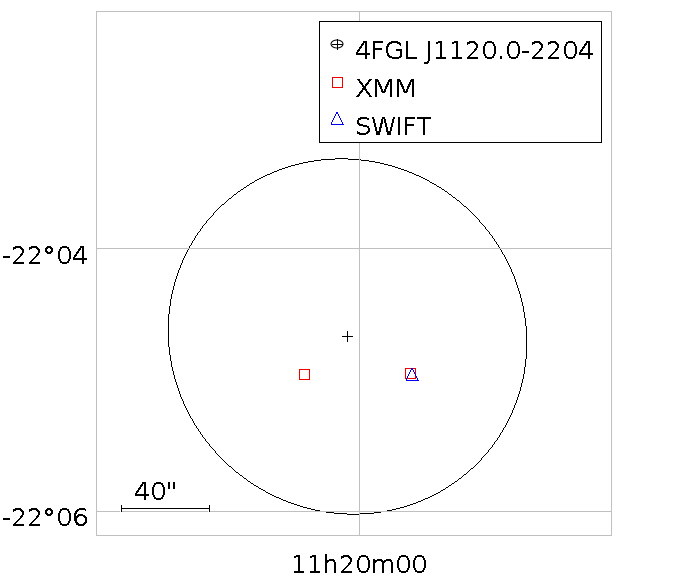}}\\
{\includegraphics[width=0.32\textwidth]{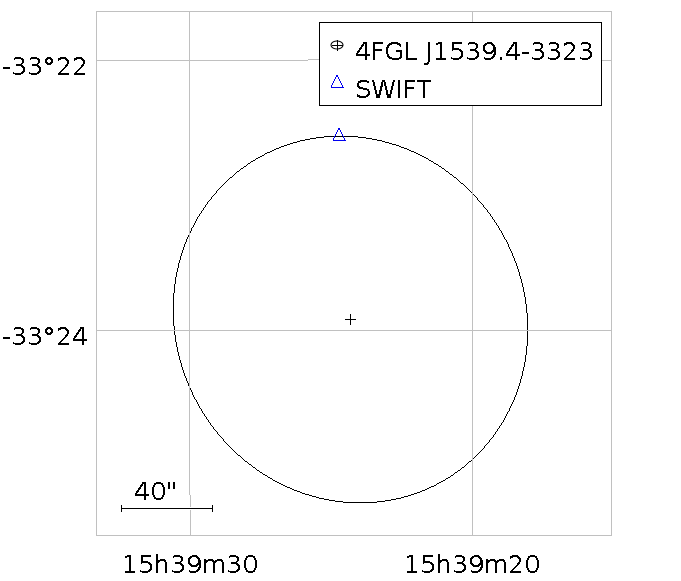}}&
{\includegraphics[width=0.32\textwidth]{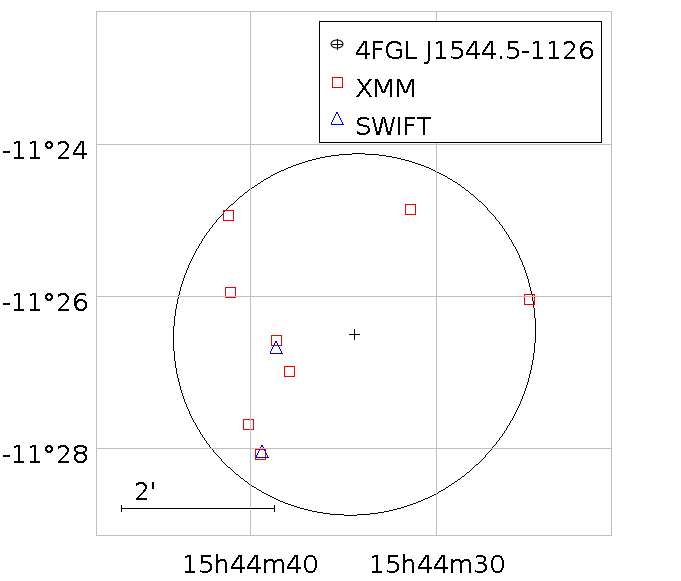}}&
{\includegraphics[width=0.32\textwidth]{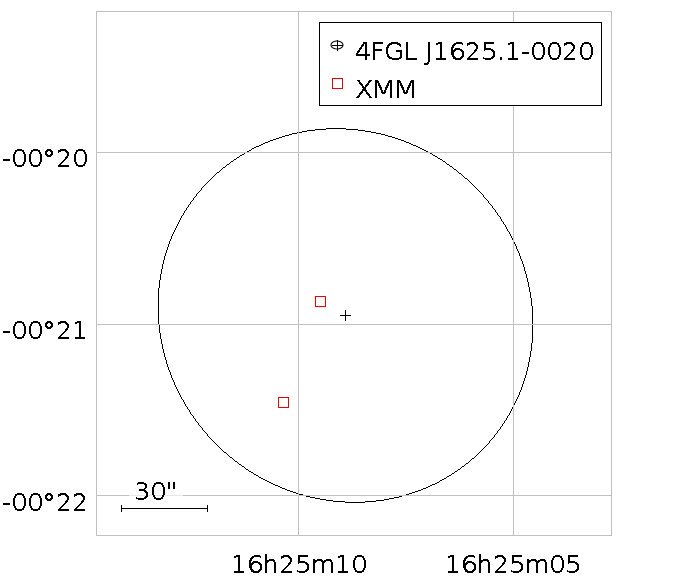}}\\
\end{tabular}
\caption{4FGL 95\% confidence error ellipses of the 23 $\gamma$-ray MSP candidates of \citet{SazParkinson2016} with X-ray coverage from {\em XMM-Newton}, {\em Swift} or {\em Chandra}. X-ray sources from the corresponding X-ray catalogues are overplotted and marked with different symbols and colours (see legenda).}
\label{fig:coverage}
\end{figure*}
\renewcommand{\thefigure}{2b}
\begin{figure*}
\centering
\begin{tabular}{ccc}
{\includegraphics[width=0.32\textwidth]{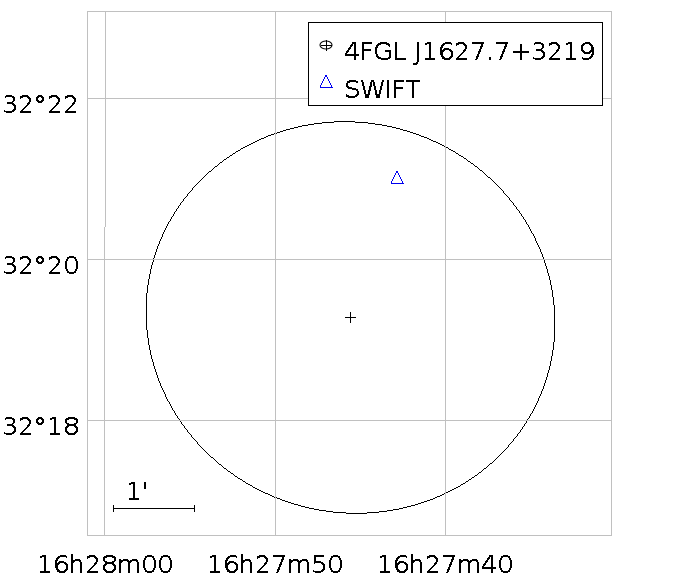}} &
{\includegraphics[width=0.32\textwidth]{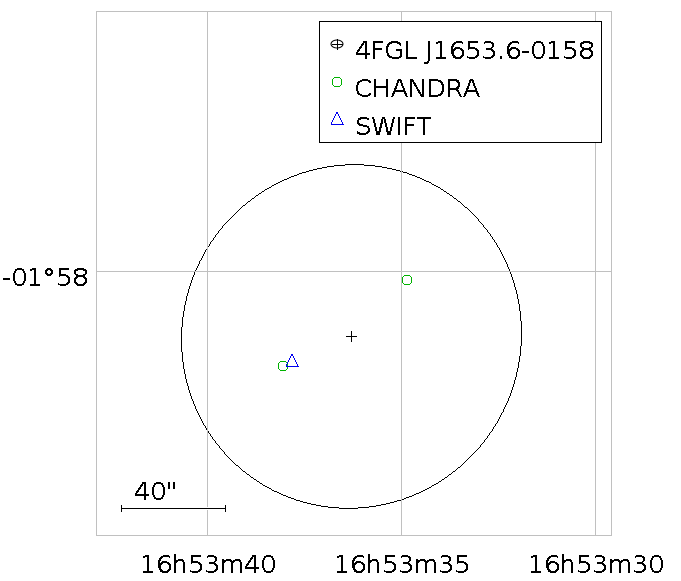}}&
{\includegraphics[width=0.32\textwidth]{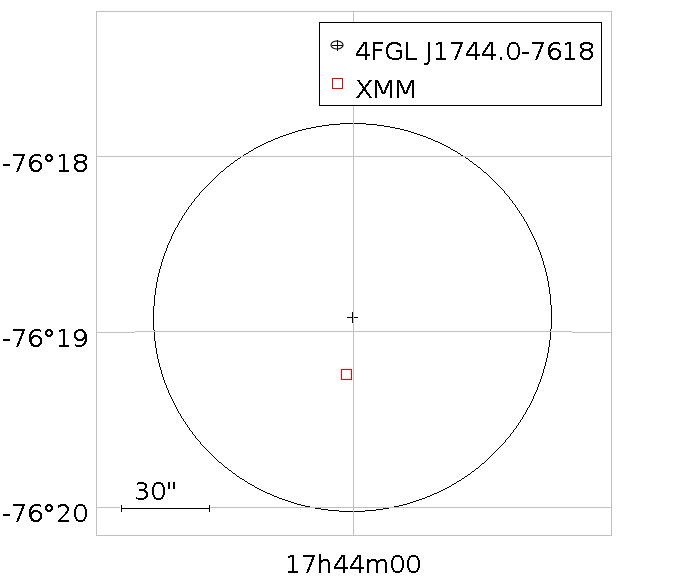}}\\
{\includegraphics[width=0.32\textwidth]{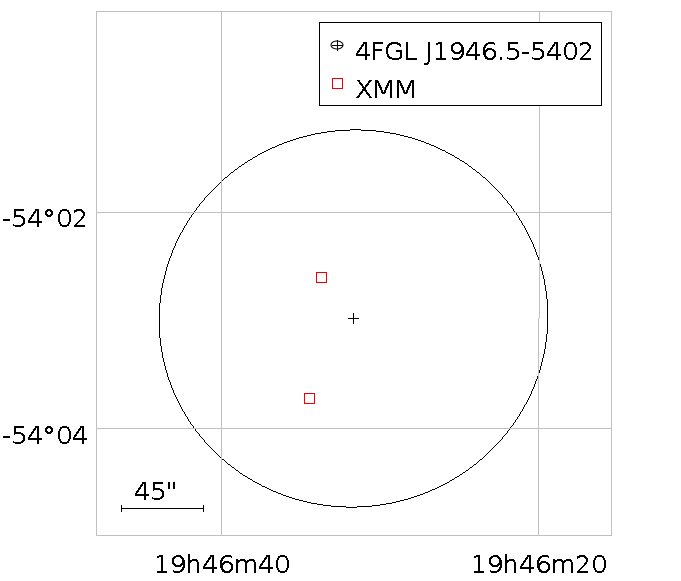}}&
{\includegraphics[width=0.32\textwidth]{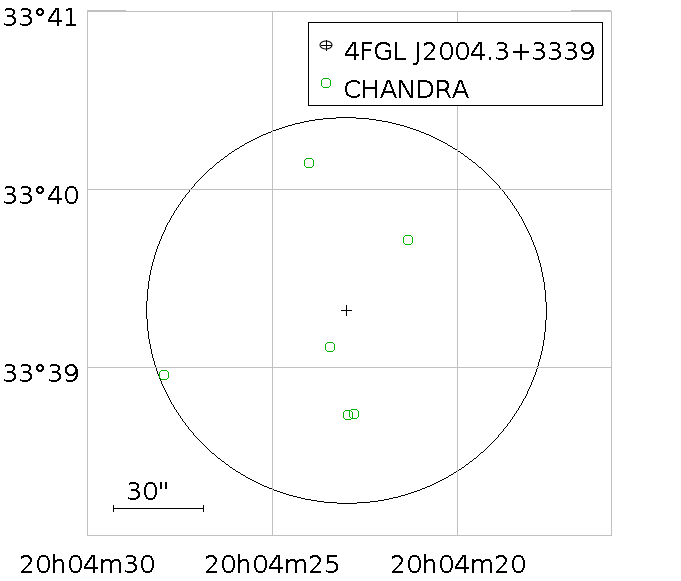}}&
{\includegraphics[width=0.32\textwidth]{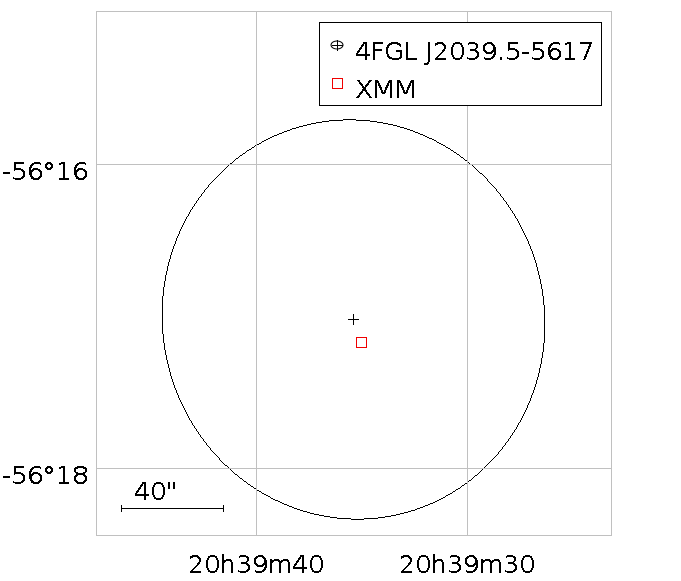}}\\
{\includegraphics[width=0.32\textwidth]{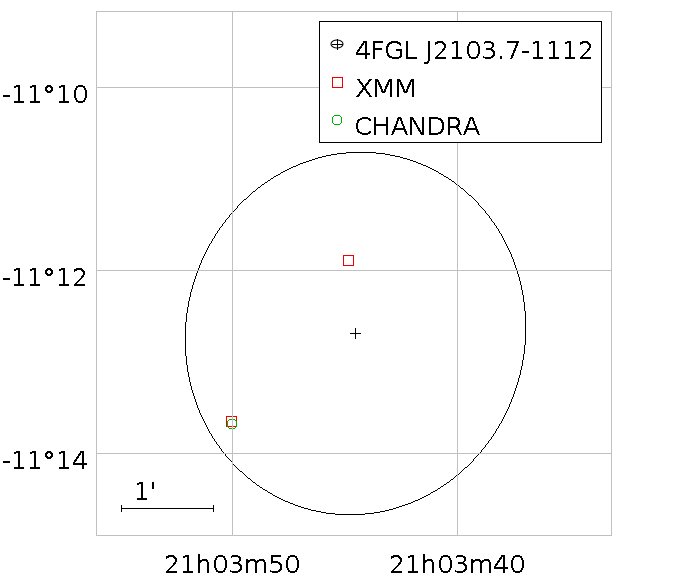}}&
{\includegraphics[width=0.32\textwidth]{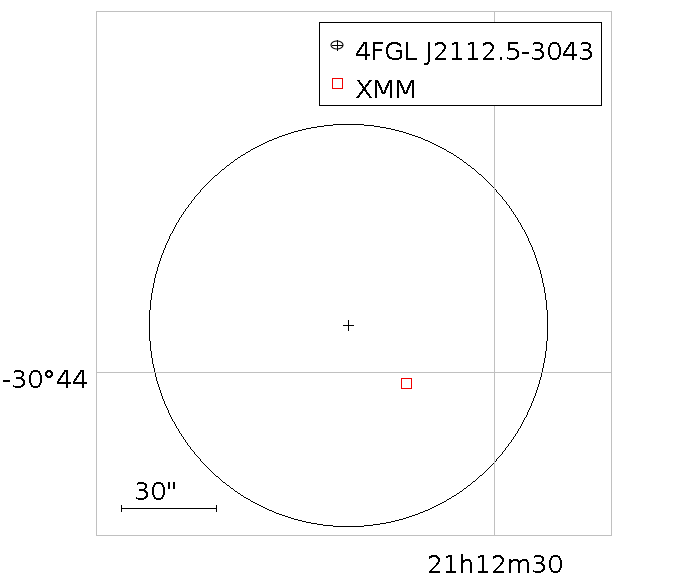}}&
{\includegraphics[width=0.32\textwidth]{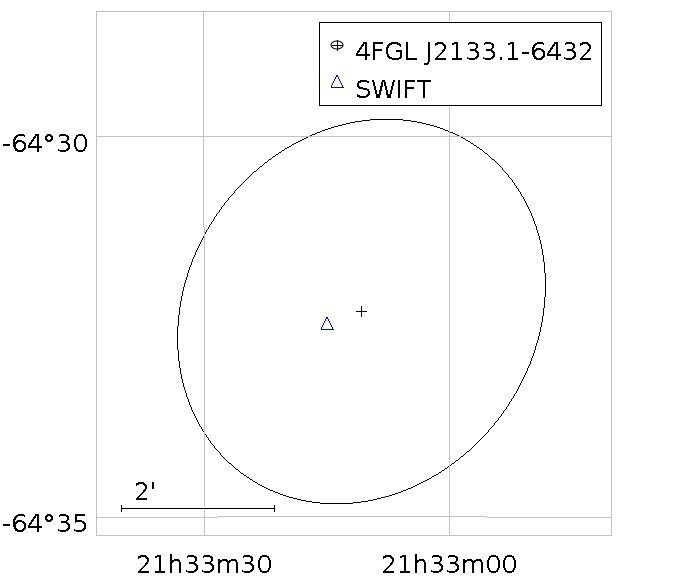}}\\
{\includegraphics[width=0.32\textwidth]{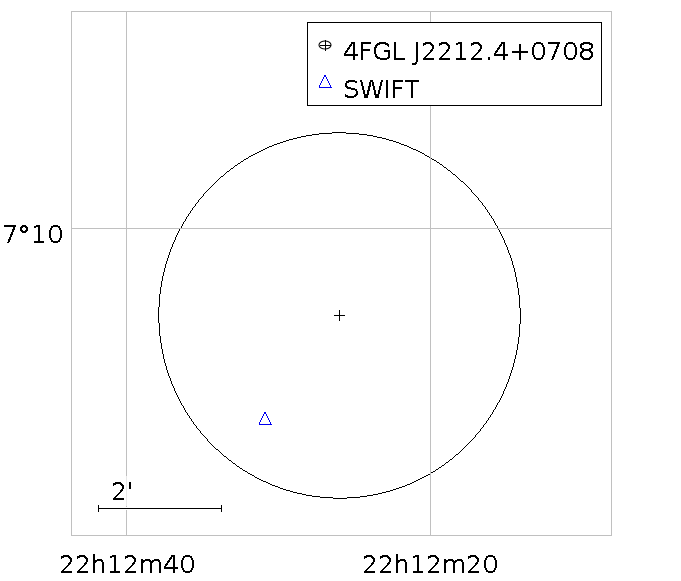}}&
{\includegraphics[width=0.32\textwidth]{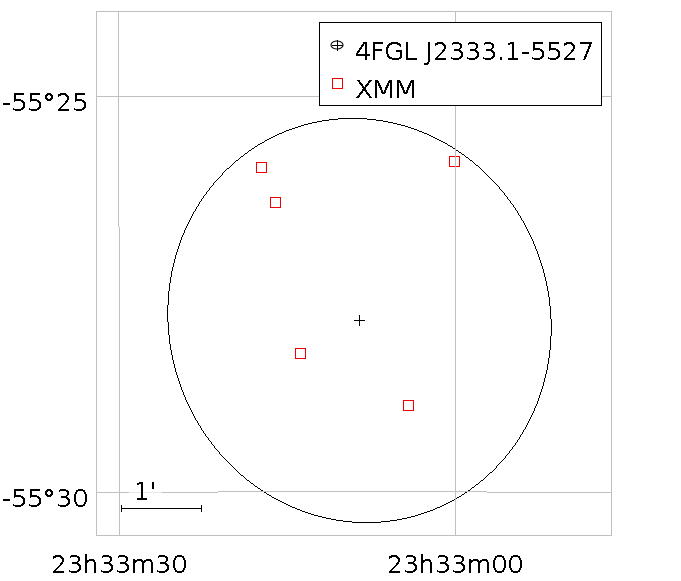}} & 
\end{tabular}
\caption{Same as in Fig.\, \ref{fig:coverage} (continued).}
\label{fig:coverage2}
\end{figure*}

\setcounter{figure}{2}
\renewcommand{\thefigure}{\arabic{figure}}
For the $\gamma$-ray sources with X-ray coverage from more than one X-ray satellite we checked how many X-ray candidate counterparts have been detected by more than one satellite. This we did by matching the X-ray source coordinates using a radius of 5\arcsec\ to account for the absolute astrometry accuracy in the detector focal plane of each X-ray satellite. For {\em XMM-Newton} the median value is generally $1\farcs5$\footnote{{\tt Calibration technical note XMM-SOC-CAL-TN-0018}},
for {\em Swift} it is $5\farcs5$ \citep[90\% confidence;][]{Evans2013}, while for {\em Chandra} it is $0\farcs8$ (90\% confidence), up to a maximum of $2\arcsec$ for sources observed at large off-axis angles\footnote{{\tt https://cxc.cfa.harvard.edu/cal/ASPECT/celmon/}}. 
The coordinate match implies 64 unique candidate X-ray counterparts across all the 23 $\gamma$-ray sources in Table\, \ref{tab:my-table}.

Of course, given the sparse X-ray coverage of the $\gamma$-ray error ellipses (Fig.\,\ref{fig:coverage}, \ref{fig:coverage2}) we cannot rule out that either actual X-ray counterparts to our  $\gamma$-ray sources are missed in this selection owing to the partial X-ray coverage of the error ellipses of these 23 $\gamma$-ray sources or that others lurk among  the remaining 25 $\gamma$-ray sources with no X-ray coverage at all.  In both cases, identifying candidate MSPs without an X-ray footprint by running blind periodicity searches of the several hundreds optical sources in each of the  $\gamma$-ray error ellipses is beyond the goals of this work.  These sources will be reconsidered for further investigations whenever adequate X-ray coverage is available.

The time coverage of the $\gamma$-ray source error ellipses is also different for the different X-ray catalogues (Fig.\, \ref{fig:histocover}). In particular, for most {\em Swift} sources the total  exposure time (i.e. integrated over all observations) is below $\sim$ 10 ks whereas for most {\em Chandra} sources it is below $\sim$ 50 ks. For about half of the {\em XMM-Newton} sources, however, the exposure time peaks at $\sim$ 140 and $\sim$ 180 ks, whereas for the rest it is mostly below $\sim$50 ks, and for about one third of the  {\em Chandra} sources it exceeds $\sim$400 ks. Therefore, apart from the last case, the longest exposures of the $\gamma$-ray source fields are achieved for the {\em XMM-Newton} sources.

\begin{figure}
{\includegraphics[width=0.52\textwidth]{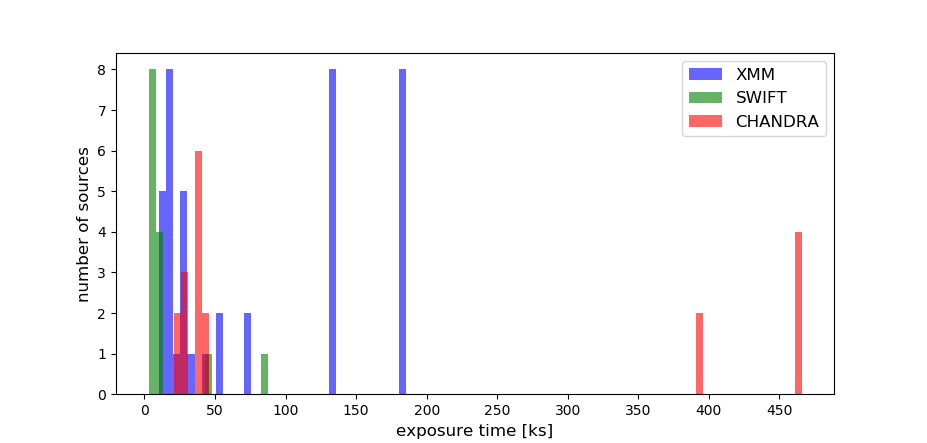}}
\caption{Histogram of the total integration time for all sources detected by {\em XMM-Newton} (41), {\em Swift} (14) and {\em Chandra} (19) in the error ellipses of the 23 $\gamma$-ray sources with X-ray coverage (Table\, \ref{tab:my-table}).}
\label{fig:histocover}
\end{figure}

Ten of the 23 $\gamma$-ray sources in Table\, \ref{tab:my-table} have already a confirmed/proposed identification: 4FGL\, J0359.4+5414 is identified with an isolated young pulsar \citep[PSR\, J0359+5414;][]{Clark2017}, 4FGL\, J1946.5-5402 with a binary MSP \citep[PSR\, J1946$-$5403;][]{Camilo2015},  4FGL\, J1035.4$-$6720 and 4FGL\, J1744.0$-$7618 with isolated MSPs \citep[PSR\, J1035$-$6720 and PSR\, J1744$-$76194;][]{Clark2018}, and they all have a pulsar association in the 4FGL (see Sectn.\ \ref{subsec:coo}). Furthermore, 4FGL\, J1544.5$-$1126 is  a candidate transitional MSP \citep{Bogdanov2015}, whereas 4FGL\, J0523.3$-$2527, 4FGL\, J0838.7$-$2827, 4FGL\, J0955.3$-$3949, and 4FGL\, J2039.5$-$5617 are candidate RBs 

\citep{Strader2014, Halpern2017a, Li2018, Salvetti2017}, and 4FGL\, J1653.6$-$0158 is a candidate BW \citep{Romani2014}, where from here on  we use the term "candidate" to refer to those sources for which the radio/$\gamma$-ray pulsation evidence has not been obtained yet.
In particular, four of the seven $\gamma$-ray sources out of the original 48 in our starting sample which have been classified as pulsars in the 4FGL catalogue (see Sectn.\, \ref{subsec:coo}) are recovered in Table\, \ref{tab:my-table} (PSR\, J0359+5414, PSR\, J1946$-$5403, PSR\, J1035$-$6720 and PSR\, J1744$-$76194).  The remaining three (PSR\, J0318+0253, PSR\, J1528$-$5838, and PSR\, J1641$-$5317) are not included in  Table\, \ref{tab:my-table} since they have no X-ray counterpart yet. Indeed, they have all been observed by {\em Swift} but the exposure time was too short (a few ks) to allow their X-ray detection. PSR\, J1528$-$5838 was also observed by {\em XMM-Newton} on August 28th 2019 with an exposure time of 18.8 ks but the data cannot be obviously included in 3XMM/DR8.

\subsection{Cross-correlation with optical catalogues}
\label{subsec:ocorr}

As a next step, we looked for candidate optical counterparts to the X-ray sources detected in the  23 candidate $\gamma$-ray source fields. Among them, we then selected those for which we found evidence of a periodic optical flux modulation of less than 1 d (Sectn.\, \ref{subsec:oper}) since this is the clear signature of the kind of binary systems we are looking for, associated with the tidally distorted and heated companion star surface (Sectn.\,\ref{subsec:strat}).
For each $\gamma$-ray source field, we performed the cross-match between the associated X-ray source list and the most recent catalogue release from the four different multi-epoch optical sky surveys discussed in Sectn.\,\ref{subsec:strat} (Catalina, PTF, ZTF, PanSTARRS) using a matching radius of $5\arcsec$. The choice of a radius of $5\arcsec$ is justified to account for the accuracy on the absolute coordinates of the X-ray sources (Sectn.\ \ref{subsec:xcorr}), which dominates over that of the optical catalogues, which is of the order of 0\farcs1--0\farcs5.  
As a safe measure against possible fake detections, we visually verified the matches directly on the optical images.

Out of the 23 $\gamma$-ray sources with candidate X-ray counterparts, 17 have also possible optical counterparts to the X-ray sources  in at least one of the selected multi-epoch surveys.   The results of the cross-match of the X-ray sources detected in the error ellipses of the 17 $\gamma$-ray sources with the selected optical catalogues are summarised in the last four columns of Table\, \ref{tab:my-table}. Like we did for the candidate X-ray counterparts (Sectn.\, \ref{subsec:xcorr}) we identified candidate optical counterparts detected across different surveys based upon their coordinates. Following our strategy, we will focus our joint X-ray/optical analysis on these 17 $\gamma$-ray sources, which passed the second screening and, thus, represent our primary working sample. The search for periodic modulations in the flux of the candidate optical counterparts to these 17 $\gamma$-ray sources is presented in Sectn.\, \ref{subsec:oper}. 

Similarly to what we discussed in Sectn.\, \ref{subsec:xcorr}, we are aware of the risk of missing actual X-ray counterparts to our candidate $\gamma$-ray sources among the six for which we have no associated optical counterpart. Indeed, their identification based on the search for orbital periodicity on the X-ray data alone is not straightforward given the average duration of the single X-ray observations  compared to that of the expected orbital periods and the serendipitous multi-epoch coverage, which for 3XMM/DR8 are $\approx24$ ks and 1--5 epochs, respectively. Nonetheless,  also spurred by the case of 3FGL\, 2039.6$-$5618 for which the orbital periodicity was firstly discovered in the X-rays \citep{Salvetti2015}, we carried out a periodicity search for all the candidate X-ray counterparts associated with these six $\gamma$-ray sources, as well as for those associated with the  17 $\gamma$-ray sources which have candidate optical counterparts (see, Sectn.\, \ref{subsec:xper}). We note that a direct search for X-ray pulsations is not possible since the observations the X-ray catalogues are built upon have not been acquired in timing mode, hence they have no adequate time resolution for periodicity searches on ms time scales.

\subsection{Optical periodicity analysis}
\label{subsec:oper}

To search for an orbital periodicity of the candidate optical counterparts we used 
an off-line computation tool based on the Lomb-Scargle (LS) periodogram algorithm.
Since the data for most of the candidate optical counterparts are unevenly-sampled in time, the LS periodogram is a suitable algorithm to use.  We used the {\sc Astropy} implementation of the LS algorithm, which is based on the code presented in \citet{VanderPlas2012} and \citet{VanderPlas2015}. 
In each periodogram, we computed the peak significance level following a procedure similar to that explained in the work of \cite{Suveges}.
This is based on the combination of both non-parametric bootstrap resampling, which allows one to reproduce the empirical distribution of the periodogram peaks, and extreme-value models which provide asymptotically valid models for the tails of more continuous distributions.
For each optical candidate counterpart, we created 1000  bootstrap repetitions of the original time series extracted from the multi-epoch sky surveys preserving the epochs of observation and replacing the object magnitudes in each observation with values randomly chosen with equal probabilities from the original data set. For each bootstrap we computed the corresponding periodogram and extracted the highest peak. The empirical distribution of the highest peaks obtained from the bootstrap was modeled on a generalized extreme-value distribution from which we computed the level corresponding to a false alarm probability of 0.01 (99\% significance threshold). Only peaks above this significance threshold were associated with a candidate periodicity in the time series.
In our analysis of the LS periodograms we carefully verified the presence of spurious peaks that could  be attributed to aliases caused by the cadence of the observations when taken at $\approx$ 1 day/cycles and around the same time in the night. 

Among the 23 $\gamma$-ray sources with possible X-ray counterparts only 17 had at least one of the X-ray sources with an associated optical counterpart (Sectn.\, \ref{subsec:ocorr}).
Among them, only seven confirmed/candidate $\gamma$-ray MSPs show evidence of orbital periodicity in at least one of their associated optical counterparts from the survey data. The results are summarised in Table\,\ref{tab:periodi}, where we report the associated optical counterparts from the different surveys and the computed period.  Four of these MSP candidates have already been classified as actual binary MSPs (BWs/RBs or other types) in previous studies and the orbital periods measured from their optical counterparts are reported in the literature, with our results in agreement with them.
Like we already stated in Sectn.\, \ref{subsec:coo}, we did not exclude these sources from our optical periodicity analysis  because they provide an essential test to establish the validity of our method and to ensure the reliability of the results obtained for the  remaining three $\gamma$-ray MSP candidates which have not been identified yet. 

The small number of optical counterparts with evidence of orbital periodicity can be naturally explained by observation biases.
Firstly, the computation of the orbital periodicity was carried out successfully only for those candidate optical counterparts for which the number of observations  was high enough to guarantee a statistically significant set of data for the LS periodogram computation. In particular, we computed the periodogram only for sources with at least 50 observations available. Secondly, not all surveys could be exploited at the same level. For instance, Catalina, PTF and ZTF explore the sky in one specific optical band, whereas PanSTARRS observes in five different bands and the number of observations per band is uneven.  In this case, the criteria for the periodogram computation was to select the band that maximises the number of observations available, which might have ended up not being always adequate. Moreover, the time span covered by the four surveys is different, which means that the number of observations that they have collected over the years is also different. Indeed, for the seven sources with periodically-modulated candidate optical counterparts, the period was mainly computed from the Catalina data and only in one case also with the PTF data, even though the same object  was detected in more than one survey. This is because Catalina and PTF are the surveys which have been running for the longest time among the four that we used.
Finally, even when enough observations were available the survey cadence was not always suitable for the investigation of a periodicity of less than about 1 day. 

\subsection{X-ray periodicity/variability analysis}
\label{subsec:xper}

Since for both RBs and BWs we expect that the X-ray flux is modulated at the orbital period of the binary system, owing to the emission from the intra-binary shock, 
we performed a systematic periodicity search in the X-ray data, targeted at periods smaller than 1 d. In principle, one should skip this step for the three $\gamma$-ray sources identified either as isolated young pulsars or isolated MSPs, i.e. 4FGL\, J0359.4+5414, 4FGL\, J1035.4$-$6720, 4FGL\, J1744.0$-$7618 (Table\, \ref{tab:my-table}). In these cases, however, an eventual chance coincidence with a periodically-modulated X-ray source would help to assess the validity of this method to identify candidate BWs and RBs. 
We focused our periodicity analysis on the {\em XMM-Newton} data, which provide most of the X-ray candidate identifications (Sectn.\,\ref{subsec:xcorr}). Furthermore, for the analysis of these data we could capitalise on the automatic tools developed within the {\em EXTraS} project\footnote{\tt http://www.extras-fp7.eu/} \citep{DeLuca2016}. There are  41 X-ray sources in 3XMM/DR8 associated with  13 $\gamma$-ray sources (Table\, \ref{tab:my-table}), including those with or without a likely optical counterpart, and we carried out the periodicity analysis for all of them. 

At the same time, we carried out a general search for long-term X-ray flux variability in the {\em XMM-Newton} data, which might reflect the transition from rotation-powered to accretion-powered states, like in transitional MSPs. Finally, since RBs also feature flaring activity \citep{Halpern2017a} we also looked for X-ray flares in the {\em XMM-Newton} observations of our $\gamma$-ray sources. Depending on the flare duration and intensity and on the persistent flux level, sources of X-ray flares can be detected only in some parts of an observation and might not appear in the 3XMM/DR8 catalogue, where the source detection is run on the whole time-integrated observation. Therefore, such sources might not be listed in Table\, \ref{tab:my-table}. In both cases, we used the tools specifically developed within the {\em EXTraS} projects.

\section{Discussion of Results}
\label{subsec:results}
Starting from the sample of the 48 $\gamma$-ray MSP candidates of \citet{SazParkinson2016}, out of the 23 with possible X-ray counterparts we extracted 17 $\gamma$-ray sources with associated optical counterparts (Table\, \ref{tab:my-table}), of which only 
six feature a more or less clear evidence of  orbital periodicity (Table\, \ref{tab:periodi}) according to our LS periodogram analysis. 
Apart from the observational biases described in Sectn.\, \ref{subsec:oper}, such a small number can also be explained by the intrinsically non-homogeneous nature of our sample.  Indeed, the 48 $\gamma$-ray MSP candidates of \citet{SazParkinson2016} in principle include both isolated and binary MSPs, as discussed in Sectn.\, \ref{subsec:strat}. In the first place, we do not expect to observe isolated MSPs since neutron stars are very faint objects in the optical \citep{Mignani2011}, far below the limiting magnitudes of the optical surveys used in this work which are in the range 22--23. Furthermore, most binary MSPs have WD companions. Because of their compactness, WDs are less subject to tidal distortion and less affected by irradiation from the MSP. Therefore, they are not expected to exhibit significant modulations of the optical flux along the binary system orbit. In conclusion, the restricted class of RBs/BWs remains the only promising target for our periodicity search. 

The results of our optical variability analysis for known and candidate BWs/RBs are discussed on a case by case basis in Sectn.\, \ref{subsec:oldcand} and \ref{subsec:newcand}, respectively. The status of proposed but yet unconfirmed BW/RB identifications for which we could not find either an optical counterpart or an optical periodicity is updated in Sectn.\, \ref{subsec:uncon}. Finally, the discussion of the results of our multi-wavelength variability analysis is supplemented in Sectn.\, \ref{subsec:xvar} which is focused  on the X-ray observations.

\subsection{Known BW/RB candidates}
\label{subsec:oldcand}
Out of the 
six known BW/RB and tMSP candidates (Table \ref{tab:my-table}) we could recover orbital periodicity of the optical counterparts only for the three RB candidates 4FGL\, J0523.3$-$2527, 4FGL\, J0955.3$-$3949, 4FGL\, J2039.5$-$5617 and for the BW 4FGL\, J1653.6$-$0158. For the candidate tMSP 4FGL\, J1544.5$-$1126 we could not find a periodicity in agreement with the published value, whereas for the remaining known RB candidate 4FGL\, J0838.7$-$2827 the available optical survey data did not allow us to run a periodicity search. 
For the binary MSP 4FGL\, J1946.5$-$5402, which is a possible BW candidate, the periodicity search did not provide convincing evidence of flux modulations at the known orbital period.

\begin{figure}
\centering
{\includegraphics[width=0.52\textwidth]{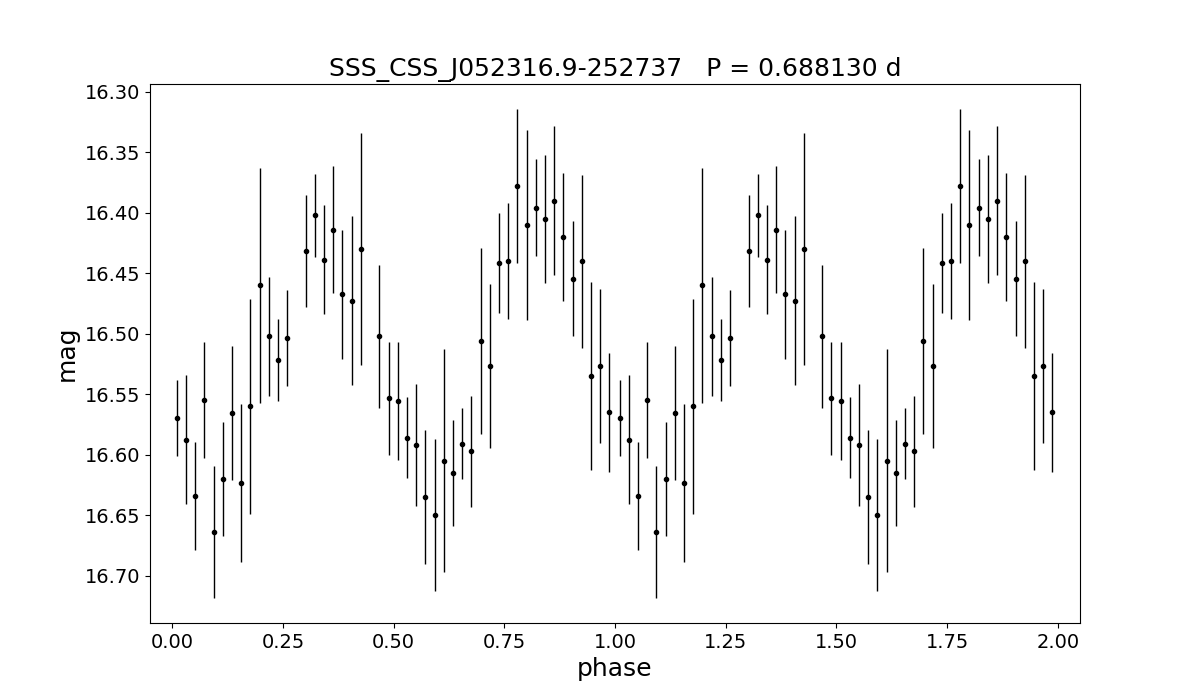}} 
\caption{
Two clear peaks are recognised in the LS periodogram at periods of 0.344065 d and its double 0.68813  d. 
Folded Catalina light curve at the longer period, which has been confirmed by optical spectroscopy \citep{Strader2014}. Two cycles are shown for clarity. A rebin of a factor of five in phase has been applied after rejection of outlier photometry measurements to better show the light curve morphology. Two slightly asymmetric maxima separated in phase by $\sim$ 0.5 are clearly recognised, as shown by \citet{Strader2014}. The absolute phase of the folded light curve is set arbitrarily.}
\label{fig:0523}
\end{figure}

\subsubsection{4FGL\, J0523.3$-$2527}
4FGL\, J0523.3$-$2527 is a candidate RB with orbital period P$_{\rm B}$=0.688134(28) d measured in the optical \citep{Strader2014}.
We found optical coverage with the Catalina (CSS and SSS), ZTF and PanSTARRS surveys but, due to the scarcity of observations in both ZTF and PanSTARRS, only the Catalina data allowed us to run the periodicity search. The companion star of the RB candidate 4FGL\, J0523.3$-$2527 is detected in both the CSS and SSS (CSS/SSS\, J052316.9$-$252737). After combining the data collected by both surveys, we obtained a sample of 244 observations from which we found a period of 0.68813  d in the LS periodogram. The corresponding folded light curve is shown in (Fig.\,  \ref{fig:0523}), 
which is in agreement with that originally measured by \citet{Strader2014} also from the CSS/SSS data and already confirmed by \citet{Salvetti2015}.  We also find an alias at half the above reported period in the periodogram. 
This was also found in the analysis of \citet{Strader2014} and ruled out as the actual period of the binary system because it was in disagreement with the value measured independently from the radial velocity curve obtained from optical spectroscopy. 

\subsubsection{4FGL\, J0838.7$-$2827}
Another of the known RB  candidates with a likely optical counterpart in our survey data is 4FGL\, J0838.7$-$2827. This source was previously studied in the work of \citet{Rea2017} where they proposed a few X-ray candidate counterparts possibly associated with the $\gamma$-ray source. One of them, 3XMM\, J083850.4$-$282757, shows variable X-ray emission, with a powerful flare \citep{Halpern2017a} similar to those observed in transitional MSPs during the sub-luminous disc state. For this reason, it is considered the most likely X-ray counterpart to 4FGL\, J0838.7$-$2827 also based upon the detection of a $\sim$ 0.21 d optical flux modulation \citep{Halpern2017b}. For this X-ray source we found an associated optical counterpart in  PanSTARRS only, but the scarcity of observations did not allow us to run the periodicity search and independently confirm the detection of the $\sim$ 0.21 d optical flux modulation \citet{Halpern2017b}. Folding the PanSTARRS data around this period does not produce evidence of periodic modulations in the  light curve and is not shown here.

\subsubsection{4FGL\, J0955.3$-$3949}
\begin{figure}
\centering
\includegraphics[width=0.52\textwidth]{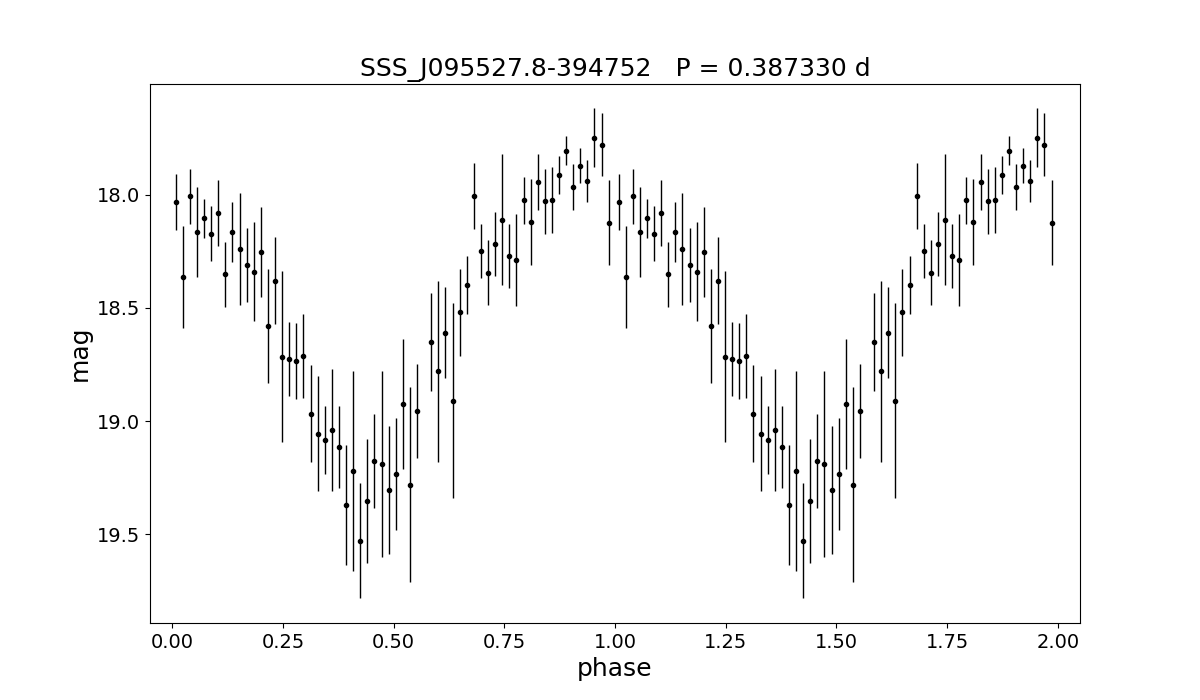} 
\caption{
Catalina light curve computed from the Catalina data of the companion star to the RB candidate 4FGL\, J0955.3$-$3949 folded at the period of the LS periodogram  main peak, 0.38733(6) d, after rebinning by a factor of three. A single broad maximum is apparent, consistent with the light curve published  in \citet{Li2018}.}
\label{fig:0955}
\end{figure}

4FGL\, J0955.3$-$3949 is a candidate RB with an orbital period P$_{\rm B}$=0.3873318(13) d, discovered from a period search at the position of its {\em Swift} candidate X-ray counterpart (1SXPS\, J165337.8$-$01583) using the Catalina data \citep{Li2018}. In this work, we recovered the X-ray source association with the same Catalina source (SSS\, 095527.8$-$394752), not detected in PTF, ZTF, and PannSTARRS, and independently searched for periodicity. We  detected a main peak in the LS periodogram 
at a period of 0.38733(6) d, which is in very good agreement with the results of \citet{Li2018}. The corresponding folded light curve is shown in (Fig.\, \ref{fig:0955}).

\begin{figure}
\centering
{\includegraphics[width=0.52\textwidth]{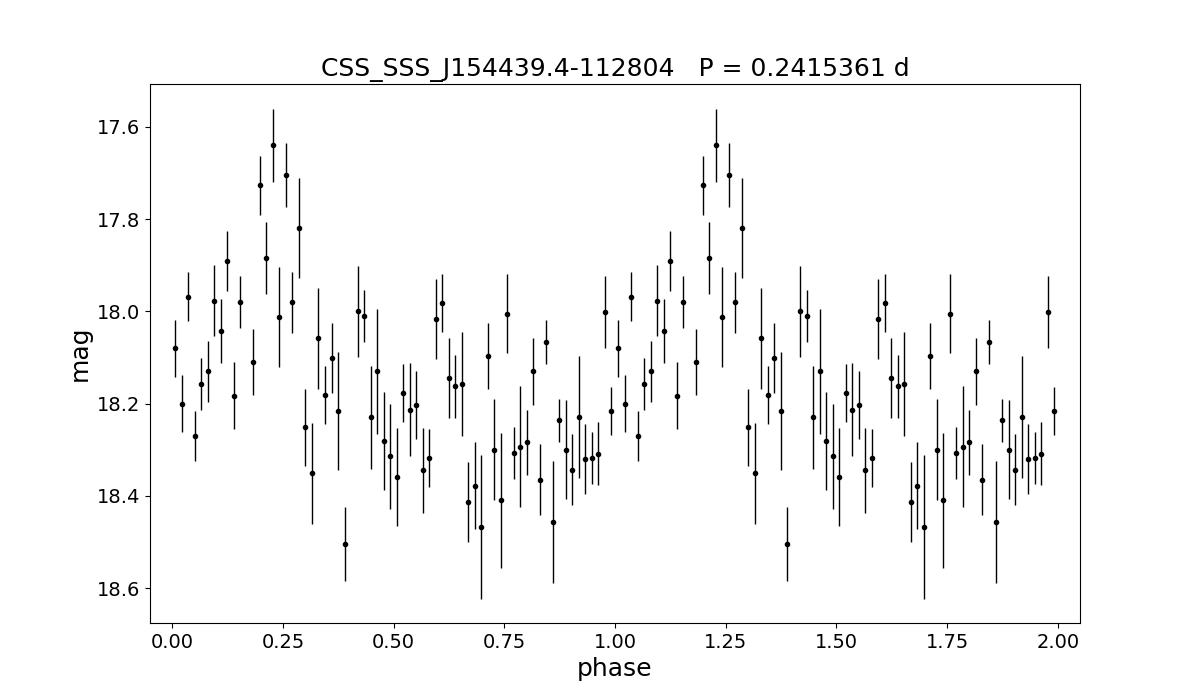}} 
\caption{Catalina light curve of the companion star to 4FGL\, J1544.5$-$1126 (CSS, SSS\, J154439.4$-$112804) folded at the orbital period P$_{\rm orb}$=0.2415361(36) d  measured from the radial velocity curve \citep{Britt2017}. A rebinning by a factor of five in phase has been applied for a better visualisation.}
\label{fig:1544}
\end{figure}

\subsubsection{4FGL\, J1544.5$-$1126}
4FGL\, J1544.5$-$1126 is a candidate transitional MSP. Radial velocity measurements obtained from optical spectroscopy of the companion star \citep{Britt2017} showed that this is a remarkably face-on binary system with inclination $i = 5^{\circ}$--$8^{\circ}$ and an orbital period  P$_{\rm B}$=0.2415361(36) d. Previous studies of the companion star light curve failed to detect a clear periodic flux modulation \citep{Bogdanov2015}. We found optical coverage from Catalina, ZTF, PanSTARRS but only for Catalina the number of observations was adequate for the periodogram computation. Both by combining the observations of the three surveys (CSS, MLS, SSS) and analysing them separately we could not find a peak in the LS periodogram corresponding to the known orbital period. 
We also folded the Catalina data at the period measured by \citet{Britt2017} but, while one can recognise a possible flux modulation with a single broad maximum (Fig.\, \ref{fig:1544}), it is not clear whether this is ascribed to a genuine orbital variability or to the folding of data affected by short-term variability characterised by a sequence of minima and maxima occurring on few hour time scales \citep{Bogdanov2015}. 
This trend cannot be recognised in the unfolded Catalina light curve owing to the coarser data sampling.
Catalina data spanning 7 years were also analysed by \citet{Bogdanov2015} who concluded that there was no significant evidence of a modulation.
Future observations should restrict the periodicity search to time intervals selected to avoid the light curve minima and maxima.

\subsubsection{4FGL\, J1653.6$-$0158}
\begin{figure}
\centering
{\includegraphics[width=0.52\textwidth]{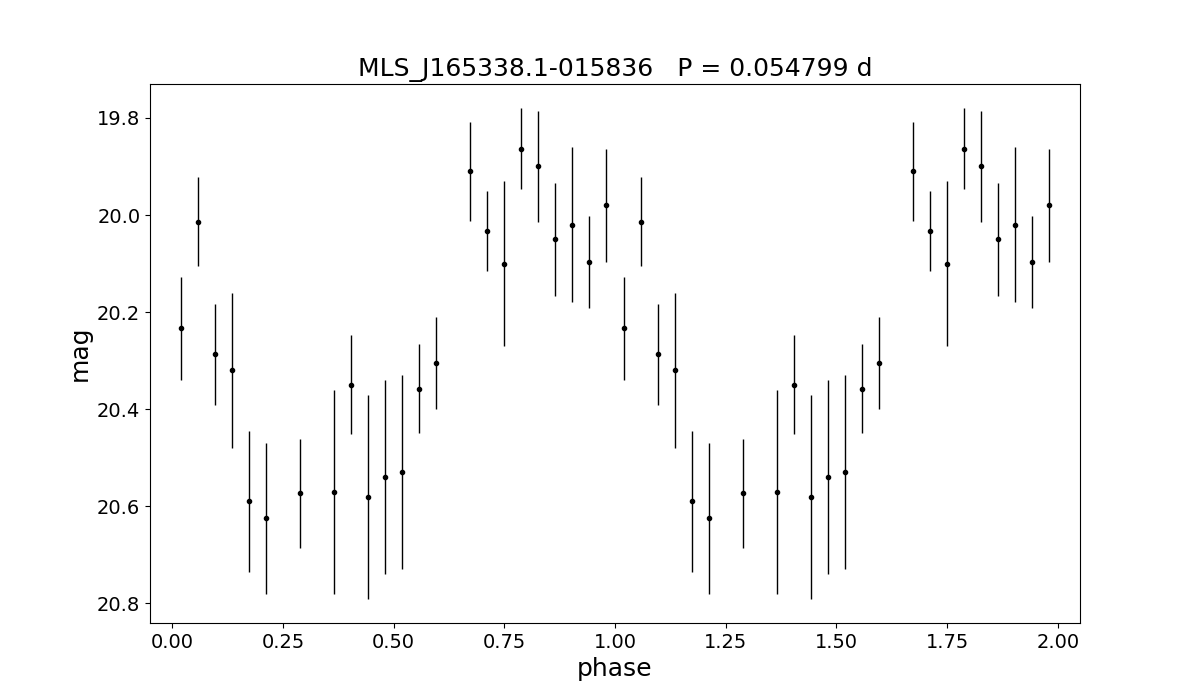}} 
\caption{
Catalina light curve from the Catalina data of the companion star to the BW candidate 4FGL\, J1653.6$-$0158. folded at the period of the periodogram  main peak, 0.054799(2) d, after rebinning by a factor of two. A single broad maximum is apparent,  consistent with the light curve published  in \citet{Romani2014}.}
\label{fig:1653}
\end{figure}

4FGL\, J1653.6$-$0158 is a candidate BW with the shortest orbital period known of P$_{\rm B}$ = 0.05194469($+$10, $-$08) d  measured by \citet{Romani2014} and obtained by combining SOAR, WIYN, and Catalina (MLS) observations of the optical counterpart. This was also studied by \citet{Salvetti2017} using the Catalina data alone and the periodicity was confirmed.  We detected the optical counterpart to 4FGL\, J1653.6$-$0158  in all the optical surveys considered in this work but only for Catalina the number of observations allowed us to run  the  periodicity search. We found a peak in the periodogram at a period of 0.054799(2) d, with the folded light curve shown in Fig.\, \ref{fig:1653}. 
The period value not in complete agreement with the value obtained by \citet{Romani2014} but is consistent with that independently obtained by \citet{Salvetti2017} using the same Catalina data (0.05479894102 d). The difference in the period determinations is probably ascribed to the sparse data and large error bars of the Catalina data with respect to the combined data set used by \citet{Romani2014}, which makes the period computation less precise.

\subsubsection{4FGL\, J1946.5-5402}
\begin{figure}
\centering
\includegraphics[width=0.52\textwidth]{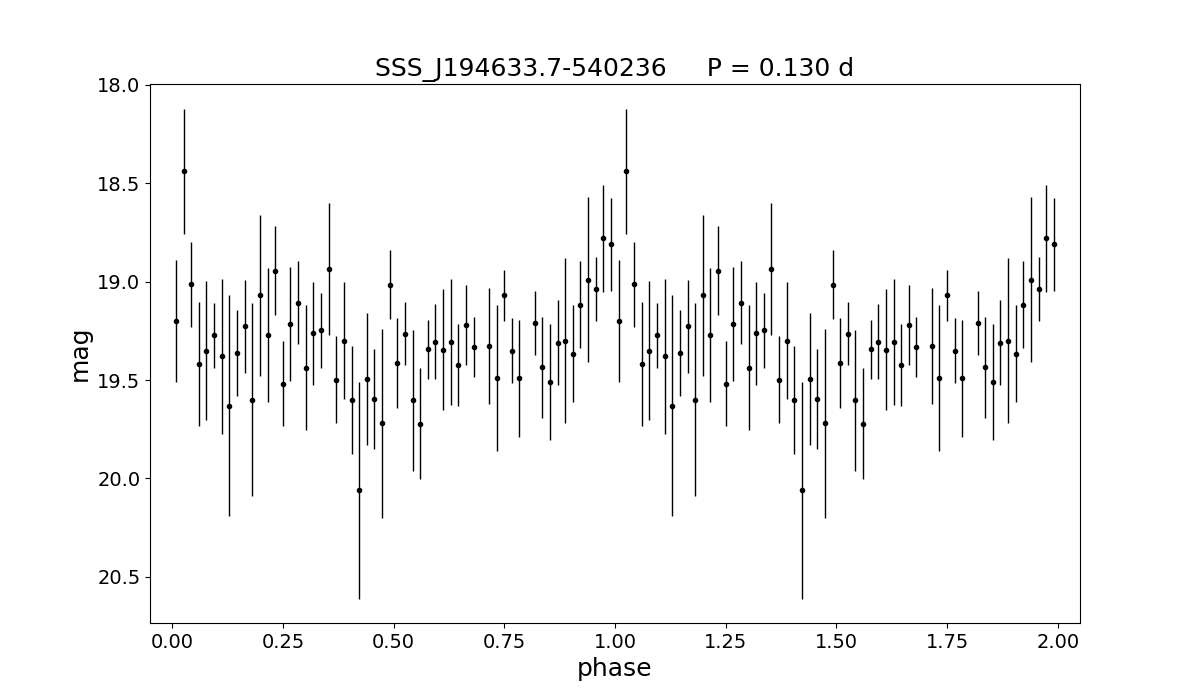}
\caption{Catalina light curve of SSS\, J194633.7$-$540236 folded at the orbital period (0.130 d) of PSR\, J1946$-$5403 measured in radio by \citet{Camilo2015}. A rebinning by a factor of five in phase has been applied for a better visualisation. }
\label{fig:1946}
\end{figure}

4FGL\, J1946.5-5402 is identified  with the binary MSP PSR\, J1946$-$5403 discovered in a Parkes radio search by \citet{Camilo2015} at a DM-derived distance of $\sim$1.15 kc \citep{Yao2017}. The system has an a orbital period of P$_{\rm B}$ = 0.130 d and, based on the lower limit on the companion star mass ($M_{\rm C} \ga 0.021 M_{\odot}$), is considered a candidate BW although no radio eclipses have been detected yet. Owing to the large uncertainty of $\sim7\arcmin$ associated with the radio position \citep{Camilo2015}, the companion star to this BW has not been identified yet and no optical periodicity has been found.  In our work we only found an object in Catalina (SSS\, J194633.7$-$540236; V=19.31),  at a position consistent with the coordinates of 3XMM\, J194633.6$-$540236, one of the two X-ray sources detected in the $\gamma$-ray error ellipse of  4FGL\, J1946.5-5402 (Fig.\, \ref{fig:coverage}). The X-ray source coordinates are within the radio position uncertainty region of PSR\, J1946$-$5403, which makes it a possible X-ray counterpart to the pulsar. Owing to the large error bars in the Catalina data of SSS\, J194633.7$-$540236, the periodogram did not reveal any significant peak at a period consistent with that of the PSR\, J1946$-$5403 orbit (0.130 d). 
Folding the Catalina data around this value (Fig.\, \ref{fig:1946}) only reveals a weak evidence of modulation which should be investigated through future observations.
Therefore, based on the optical data alone we cannot claim an association of 3XMM\, JJ194633.6$-$540236/SSS\, J194633.7$-$540236 with the pulsar, whose companion star still remains unidentified. An improvement of the pulsar radio position and a better determination of the binary period would help the identification process.

\begin{figure}
\centering
{\includegraphics[width=0.52\textwidth]{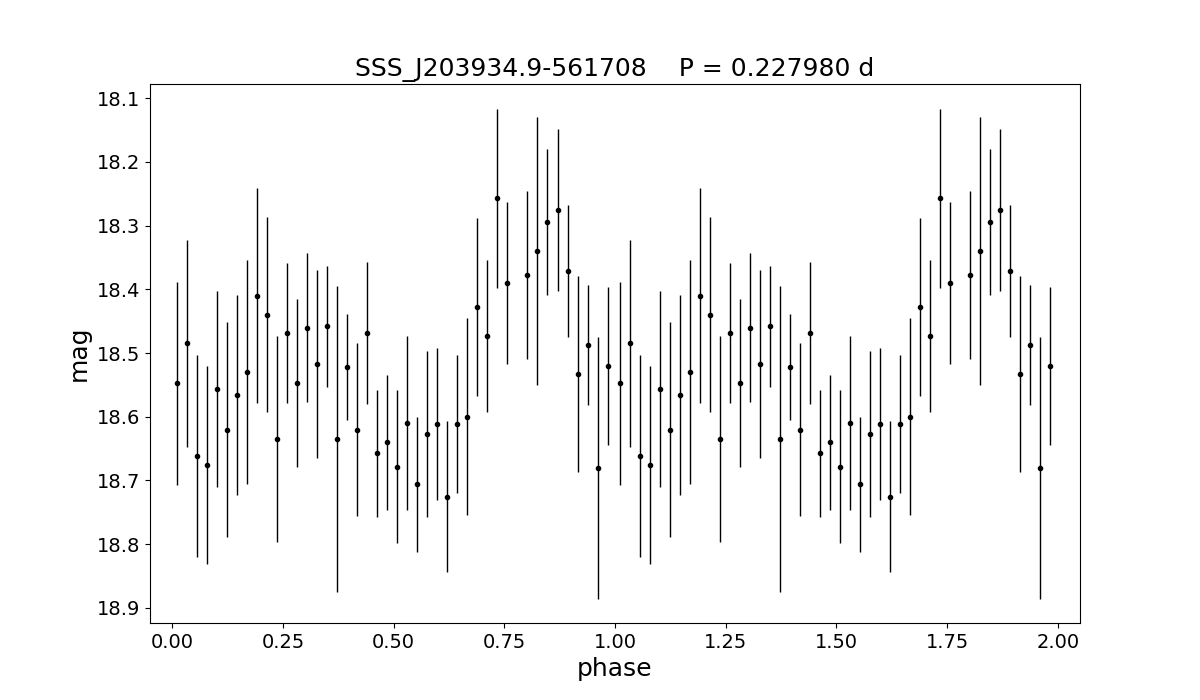}} 
\caption{
Catalina light curve of the RB candidate 4FGL\, J2039.5$-$5617 \citep{Salvetti2015} folded at the LS peak period of 0.22798(3)  d, which we measured in the Catalina data for the first time, after applying a phase rebinning by a factor of five. Two asymmetric maxima separated in phase by $\sim$ 0.5 are clearly recognised, consistent with what observed in the better signal--to--noise optical light curve obtained from the GROND data \citep{Salvetti2015}.}
\label{fig:2039}
\end{figure}

\begin{figure*}
\centering
\begin{tabular}{cc}
{\includegraphics[width=0.52\textwidth]{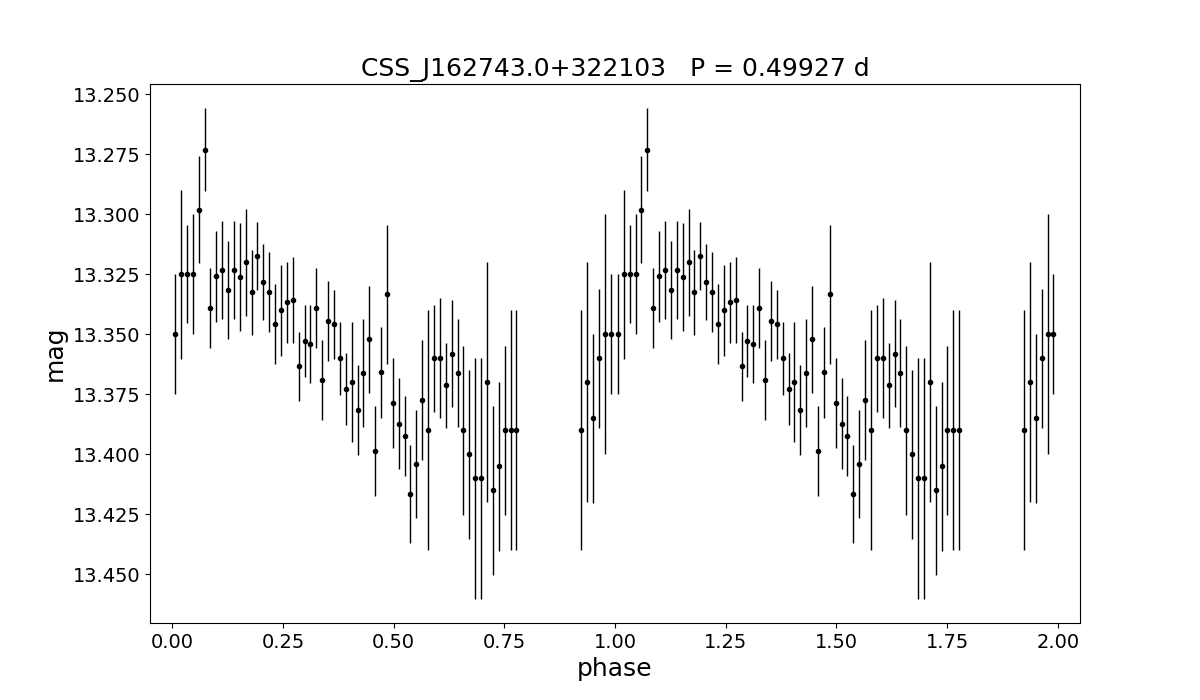}} &
{\includegraphics[width=0.52\textwidth]{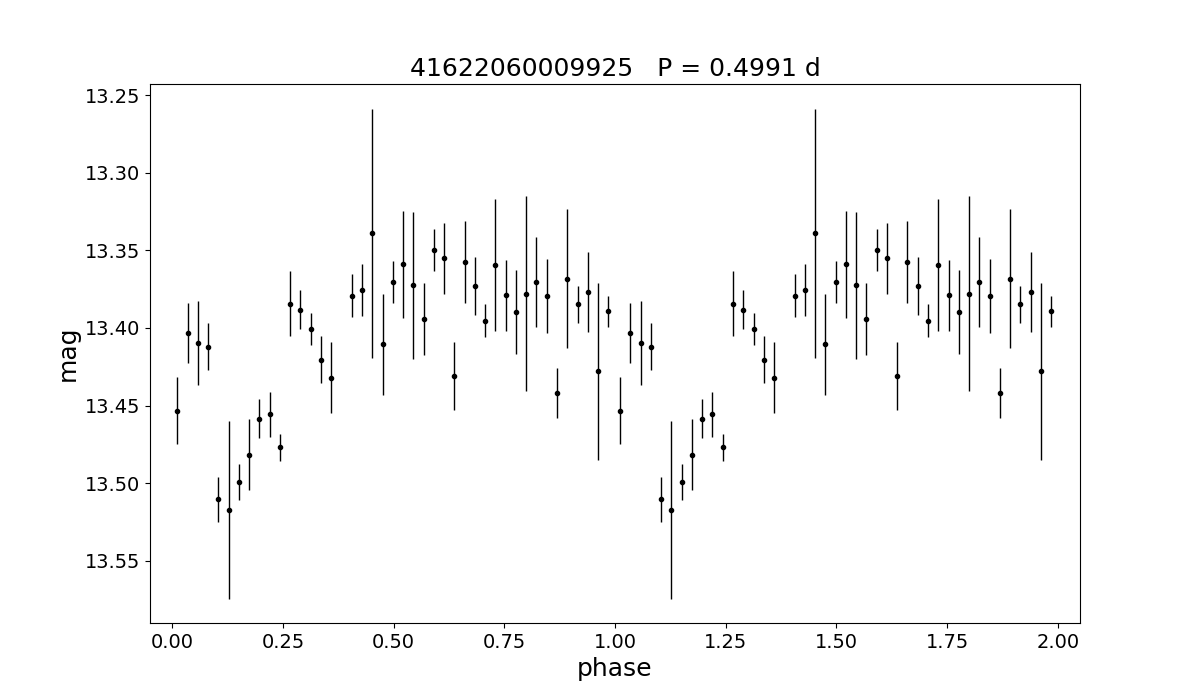}} \\
\end{tabular}
\caption{
Light curves of the candidate optical counterpart to 4FGL\ J1627.7+3219 computed from the Catalina (left) and PTF (right) data
folded at the main peak periods found in the LS periodograms, 0.49927(2) d  and 0.4991(3) d, respectively.
A rebin by a factor of three and four, respectively has been applied.}. 
\label{fig:1627}
\end{figure*}

\begin{figure}
\centering
{\includegraphics[width=0.52\textwidth]{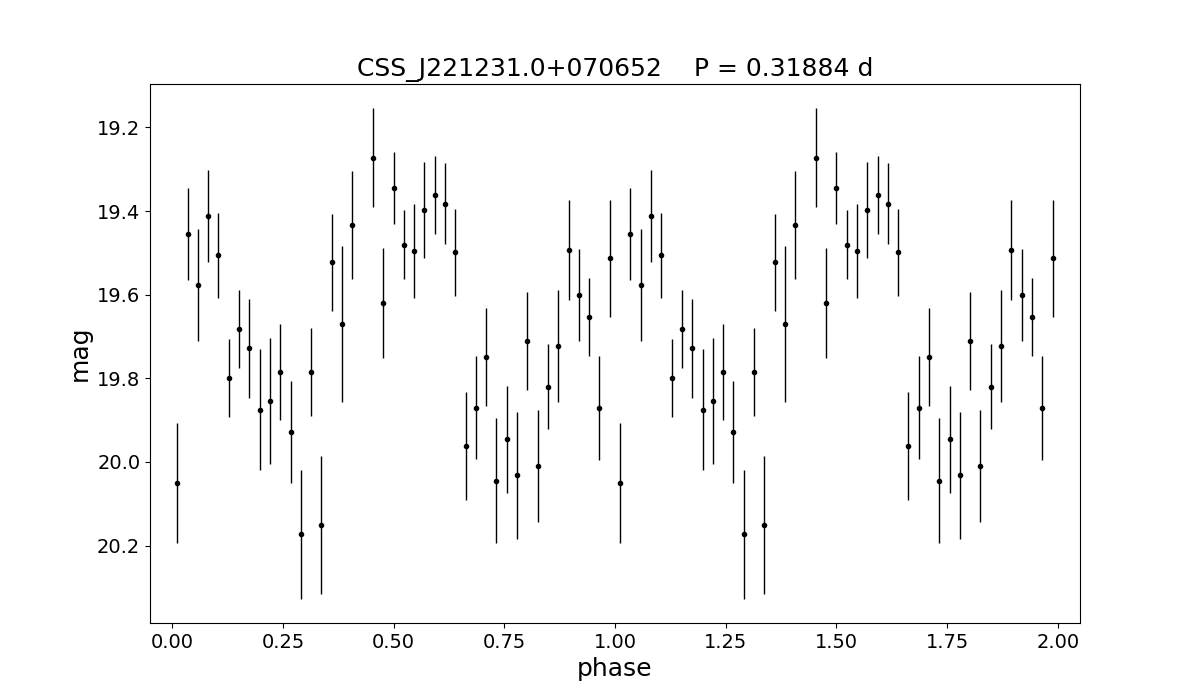}} 
\caption{Folded light curve computed from the Catalina data of the BW/RB candidate 4FGL\, J2212.4+0708 at the LS periodogram main peak of 0.31884(6)  d. }
\label{fig:2212}
\end{figure}

 \subsubsection{4FGL\, J2039.5$-$5617}
4FGL\, J2039.5$-$5617 is a candidate RB discovered by \citet{Salvetti2015} and \citet{Romani2015} for which they observed an orbital modulation in both X-rays and the optical. Using GROND data they found a period of 0.22748(43) d that coincides, within the errors, with that of the X-ray source. In this work, we found that the source was observed only by Catalina (SSS\, J203934.9$-$561708), which now provides a more extended epoch coverage (223 epochs) than available at the time when \citet{Romani2015} and \citet{Salvetti2017} carried out their periodicity search in the Catalina data.  For the first time, we now found that a periodic flux modulation is also detectable in the Catalina data, at a  period of 0.22798(3) d 
consistent with the value of the orbital period of 0.2279817(7) d obtained from radial velocity measurements by \citet{Strader2019}. As expected, the folded Catalina light curve (Fig.\, \ref{fig:2039}) is characterised by two asymmetric maxima separated in phase by $\sim$ 0.5, which are less clearly resolved in the more noisy Catalina data. 
For this reason, we could not find significant evidence of light curve evolution between the epochs of the GROND and Catalina data.

\subsection{New BW/RB candidates}
\label{subsec:newcand}
The broad agreement between our measured periods and those reported in the literature for four out of the 
six cases described in the previous section confirms the reliability of our procedure. We now applied it in the search for new BWs/RBs among the ten candidate $\gamma$-ray MSPs  with associated X-ray/optical counterparts (Table\, \ref{tab:my-table}), for which we searched for a possible optical periodicity for the first time. In particular, there are 
two candidate $\gamma$-ray MSPs in our sample,
4FGL\ J1627.7+3219
and 4FGL\, J2212.4+0708, that show peaks in the LS periodogram which are above the significance threshold and correspond to periods in the range expected for BWs/RBs.  We investigate the candidate optical periodicity of these 
two sources in the following sections. For all the remaining 
eight candidate $\gamma$-ray MSPs with an associated optical counterpart (Table \ref{tab:my-table}) we found either no significant peak in the LS periodogram or we had not enough flux measurements in any of the reference optical surveys for the periodicity search.

\subsubsection{4FGL\ J1627.7+3219}
For this candidate $\gamma$-ray MSP we found a possible X-ray counterpart detected by {\em Swift} (1SXPS\, J162742.8+322059) and an associated, quite bright (V=13.51), optical counterpart  in each of the reference optical surveys.  If this object were indeed the companion star in a BW/RB system,  the comparison with the optical properties of other BWs/RBs would presumably put it a distance of the order of a few hundred pc.  The 4FGL\ J1627.7+3219 0.1--100 GeV energy flux 
is $(3.3613 \pm 0.35810) \times 10^{-12}$ erg cm$^{-2}$ s$^{-1}$ which, for a distance of 500 pc, would correspond to a $\gamma$-ray luminosity of $\sim 10^{32}$ erg s$^{-1}$. This value is within the $\gamma$-ray luminosity range of all MSPs but a factor of 10--100 below that of BWs/RBs \citep{Hui2019}. This would imply a correspondingly larger distance for 4FGL\ J1627.7+3219, if it were indeed a BW/RB, and would argue against the association with its V=13.51 candidate optical counterpart.  That said, we investigated a possible association. Only in PTF and Catalina the number of observations was adequate to run the periodicity search. The LS periodogram computed from  the Catalina data shows a main peak at a period of 0.49927(2) d with a significance of 7 $\sigma$.
From the periodicity analysis of PTF data we found a main peak at a period 
0.4991(3) d, similar to that found with Catalina data, with a significance of 4.2 $\sigma$.  
Folding the Catalina and PTF light curves at the corresponding peak periods gives  very similar profiles, with a single broad maximum (Fig.\ref{fig:1627}).

\subsubsection{4FGL\, J2212.4+0708}
This candidate MSP has only one X-ray counterpart found with {\em Swift} (1SXPS\, J221230.8+070651) and it is associated with an optical counterpart for which we found a possible periodic modulation.  This object (V=19.7)  has been detected by all the four reference optical surveys (Table\, \ref{tab:my-table}) but, also in this case, only the data collected by Catalina cover a range of epochs large enough to allow the LS periodogram computation. We found a possible period at 0.31884(6)  d, corresponding to the main peak in the LS periodogram, which, however, is only slightly above the computed significance threshold. We folded the Catalina light curve at this period applying a rebinning of a factor five in phase. The folded optical light curve (Fig.\, \ref{fig:2212}) shows a possible modulation with two asymmetric maxima, as observed in most RB systems.  However, the low significance of the peak in the periodogram makes it difficult to determine whether this periodicity is real. Therefore, we cannot confidently affirm that 4FGL\, J2212.4+0708 is a RB candidate, although the candidate periodicity is worth investigation through follow-up observations.

\subsection{Unconfirmed BW/RB candidates}
\label{subsec:uncon}
Five of the unclassified sources in Table\, \ref{tab:my-table} (4FGL\, J0802.1$-$5612, 4FGL\, J1120.0$-$
2204, 4FGL\, J1539.4$-$3323, 4FGL\, J1625.1$-$0020, 4FGL\, J2112.5$-$3043) have been previously studied in the optical  by \citet{Salvetti2017} using both  Catalina and dedicated GROND observations and for one of them (4FGL\, J0802.1$-$5612) an orbital periodicity (0.4159 d) was proposed based on the Catalina data. Here, we extended their analysis by adding observations from the PTF, ZTF and PanSTARRS surveys, with the aim of confirming the proposed periodicity for 4FGL\, J0802.1$-$5612 and discovering a candidate periodicity for the other 
two sources. 

\subsubsection{4FGL\, J0802.1$-$5612}
The candidate optical counterpart to 4FGL\, J0802.1$-$5612 by \citet{Salvetti2017} is  SSS\, J080225.1$-$560543, associated with the X-ray source 3XMM\, J080225.3$-$560542, identified with their field source \#5. However, this candidate counterpart now falls clearly outside the revised 4FGL $\gamma$-ray error ellipse\footnote{In their work, \citet{Salvetti2017} assumed the, back-then, most recent 3FGL coordinates as a reference.} (Fig.\, \ref{fig:xgamma}) so that we cannot claim an association with 4FGL\, J0802.1$-$5612 any longer. For this reason, we did not investigate the SSS\, J080225.1$-$560543 variability in other multi-epoch optical surveys. None of the optical counterparts associated with the {\em XMM-Newton} candidate counterparts to 4FGL\, J0802.1$-$5612 (Table\, \ref{tab:my-table}) show evidence of variability in the Catalina data.  Therefore, the identification of 4FGL\, J0802.1$-$5612 as a binary MSP  remains unconfirmed. 

\subsubsection{4FGL\, J1120.0$-$2204, 4FGL\, J1539.4$-$3323, 4FGL\, J1625.1$-$0020, 4FGL\, J2112.5$-$3043}
\begin{figure*}
    \centering
    \includegraphics[width=0.8\textwidth]{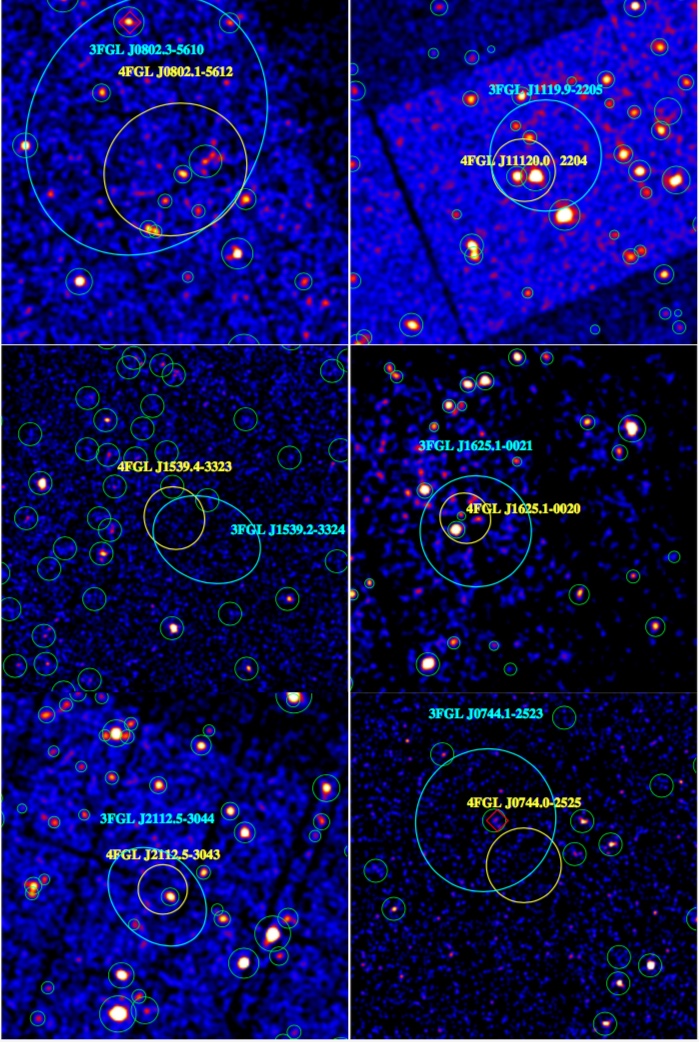}
        \caption{3FGL and the 4FGL error ellipses (cyan and yellow, respectively) for the $\gamma$-ray MSP candidates described in Sectn.\, \ref{subsec:uncon} overlaid on the X-ray image from either {\em XMM-Newton}  or {\em Swift} (4FGL\, J0744.0$-$2525 and 4FGL\, J1539.4$-$3323) smoothed with a Gaussian filter (three pixel radius). All images are $0\fdg25 \times 0\fdg25$ in size. Sources from the XMM-DR8 and 1SXPS catalogues are marked by the green circles 
    For both 4FGL\, J0802.1$-$5612 and 4FGL\, J0744$-$2525  
   the position of the candidate optical counterparts of \citet{Salvetti2017} is marked by the red diamond. 
    }
\label{fig:xgamma}
\end{figure*}

The difference between the 3FGL and 4FGL $\gamma$-ray error ellipses for 4FGL\, J1120.0$-$2204, 4FGL\, J1539.4$-$3323, 4FGL\, J1625.1$-$0020, 4FGL\, J2112.5$-$3043 is shown in Fig.\, \ref{fig:xgamma}. For 4FGL\, J1120.0$-$2204, the new $\gamma$-ray error ellipse lies within the 3FGL one and no new candidate X-ray counterparts are found in the {\em XMM-Newton} data with respect to \citet{Salvetti2017}. The optical periodicity of the two current candidate X-ray counterparts 3XMM\, J111958.3$-$22045 and 3XMM\, J112001.7$-$2204 (Table \ref{tab:my-table}), coincident with sources \#1 and \#2 of \citet{Salvetti2017}, have been searched by these authors using Catalina and GROND data but no evidence thereof was found. We re-ran the search in the Catalina as well as the ZTF and PanSTARRS data but, again, we found no evidence of optical periodicity.  For 4FGL\, J1539.4$-$3323, we found now a candidate {\em Swift} X-ray counterpart (1SXPS\,J153924.7$-$332233), whereas for 4FGL\, J2112.5$-$3043 we now found only a candidate {\em XMM-Newton} counterpart (3XMM\, J211232.1$-$304403) identified with field source \#1 of \citet{Salvetti2017}. In both cases, however, we found no associated optical counterpart in any of the reference surveys. 4FGL\, J1625.1$-$0020 has two candidate X-ray counterparts (3XMM\, J162509.4$-$002052 and 3XMM\, J162510.3$-$002127), identified with field sources \#2 and \#1 of \citet{Salvetti2017}, respectively. They both have potential PanSTARRS associations and the former has also a Catalina association, already investigated in \citet{Salvetti2017}, which, however, show no corresponding optical modulation.  Therefore, the MSP identification for these four $\gamma$-ray sources remains unconfirmed, too. 

{}

\subsubsection{4FGL\, J0744.0$-$2525}
Another MSP candidate  in the list of \citet{SazParkinson2016} for which an optical counterpart was proposed by \citet{Salvetti2017} based upon a clear  periodic modulation (0.115 d) detected in the GROND data is 4FGL\, J0744.0$-$2525. This RB candidate does not appear in Table\, \ref{tab:my-table} 
because our cross-correlations could not find an associated X-ray counterpart within the updated 4FGL $\gamma$-ray error ellipse, which is only covered by {\em Swift} observations \citep{Salvetti2017}.
Regardless of that, having an optical identification being proposed an independent assessment is in order.  We found that the candidate GROND counterpart now falls $\sim 15\arcsec$ away from the $\sim 1\farcm6$-wide 4FGL $\gamma$-ray error ellipse (Fig.\, \ref{fig:xgamma}), 
which makes the association somewhat less likely, although not strictly incompatible accounting for possible unknown systematics in the 4FGL position determination. Therefore,  we followed up on the proposed optical identification with the PTF, ZTF and PanSTARRS surveys (the field is not covered by Catalina). We found that the GROND object has been detected only once in ZTF but repeatedly in PannSTARRS. We carried out a periodicity search in this data using the same approach as described in Sectn.\,\ref{subsec:oper}. 
However, the number of observations obtained with PanSTARRS is still too small to allow the detection of a significant optical modulation.
Therefore, we cannot add information on the light curve characteristics of the candidate optical counterpart proposed by \citet{Salvetti2017} and on their possible long-term evolution. Dedicated follow-up observations, aimed at a radial velocity measurement and an X-ray detection, are needed to verify its association with 4FGL\, J0744.0$-$2525 and confirm that this $\gamma$-ray source is indeed a RB candidate.

\subsection{X-ray variable BW/RB candidates}
\label{subsec:xvar}

Like we explained in Sect,\, \ref{subsec:xper}, we carried out a systematic search for both periodic and aperiodic  X-ray variability with the {\em EXTraS} tools\footnote{See {\tt http://www.extras-fp7.eu/index.php/archive} to access the tool documentation.} for all the 41 {\em XMM-Newton} sources detected in the fields of 13 $\gamma$-ray sources (Table\, \ref{tab:my-table}). This we did regardless of an association with either known BW/RB candidates or other types of pulsars and of the association with an optical counterpart in the reference surveys, which makes our analysis bias-free.  We note that the data products from the {\em EXTraS} variability analysis available online in the {\em EXTraS} archive\footnote{\tt https://www88.lamp.le.ac.uk/extras/archive} are still based on observations included in 3XMM catalogue releases earlier than DR8, mostly DR4 (up to December 2012) and DR5 (up to December 2013), which were the reference at the time the {\em EXTraS} project was carried out (2014--2016). Therefore, a major part of this work was to run the {\em EXTraS} variability analysis off-line to extend the results to all observations included in 3XMM/DR8 (up to November 2017).

\subsubsection{Transient and aperiodic X-ray variability}
First of all, to make sure that we did not miss potential X-ray counterparts to the $\gamma$-ray MSP sources  we used the off-line {\em EXTraS} tools to search for X-ray sources with flaring activity characterised by rapid transitions from "off" to "on" states within the same {\em XMM-Newton} observation. These are sources which switch from count rates  below the detection threshold to count rates well above it, which might escape automatic source detection and do not end up in the 3XMM catalogue (see Sectn.\, \ref{subsec:xper}). After setting the detection threshold to 200 counts in the 0.3--10 keV energy band and the maximum flare duration to 10 ks, we indeed found one such flaring sources,  which is not in the 3XMM/DR8 catalogue. In particular, we discovered this source in the {\em XMM-Newton} observation (OBS ID 0112200301; 29.6 ks) of the $\gamma$-ray source 4FGL\, J0359.4+5414, now identified as the isolated young $\gamma$-ray pulsar PSR\, J0359+5414 \citep{Clark2017}, with a flare duration of $\sim$ 3.8 ks. This source, however, is at coordinates $\alpha =03^{\rm h}  58^{\rm m} 49\fs73$; $\delta  = +54^\circ 12\arcmin 54\farcs7$ and, whatever its nature, it is clearly not associated with the pulsar which is at $\alpha =03^{\rm h}  59^{\rm m} 26\fs01$; $\delta  = +54^\circ 14\arcmin 55\farcs7$ \citep{Clark2017}.  
No other "on/off" flaring source has been detected  in the {\em XMM-Newton} observations of the fields of the remaining  12 $\gamma$-ray sources.

Incidentally, we note that source 3XMM\, J035925.2+541455 lies at only 7\arcsec\ (slightly larger than our assumed matching radius, Sect.\, \ref{subsec:ocorr}) from the position of 2CXO\, J035926.0+541455 (Table\, \ref{tab:my-table}), which is the candidate X-ray counterpart to PSR\, J0359+5414 detected by \citet{Zyuzin2018} in a $\sim$460 ks {\em Chandra} observation. The source is at the centre of a 30\arcsec-long extended X-ray emission which might have not been fully resolved by {\em XMM-Newton}, affecting its accuracy on the source position determination and the match with the {\em Chandra} one. Thus, it seems most likely that the X-ray counterpart to PSR\, J0359+5414 has also been detected by {\em XMM-Newton}. Due to the shorter exposure time, however, this observation would not add more information on the X-ray emission of PSR\, J0359+5414 with respect to the {\em Chandra} one.

\begin{figure}
    \centering
    \includegraphics[width = 0.52\textwidth]{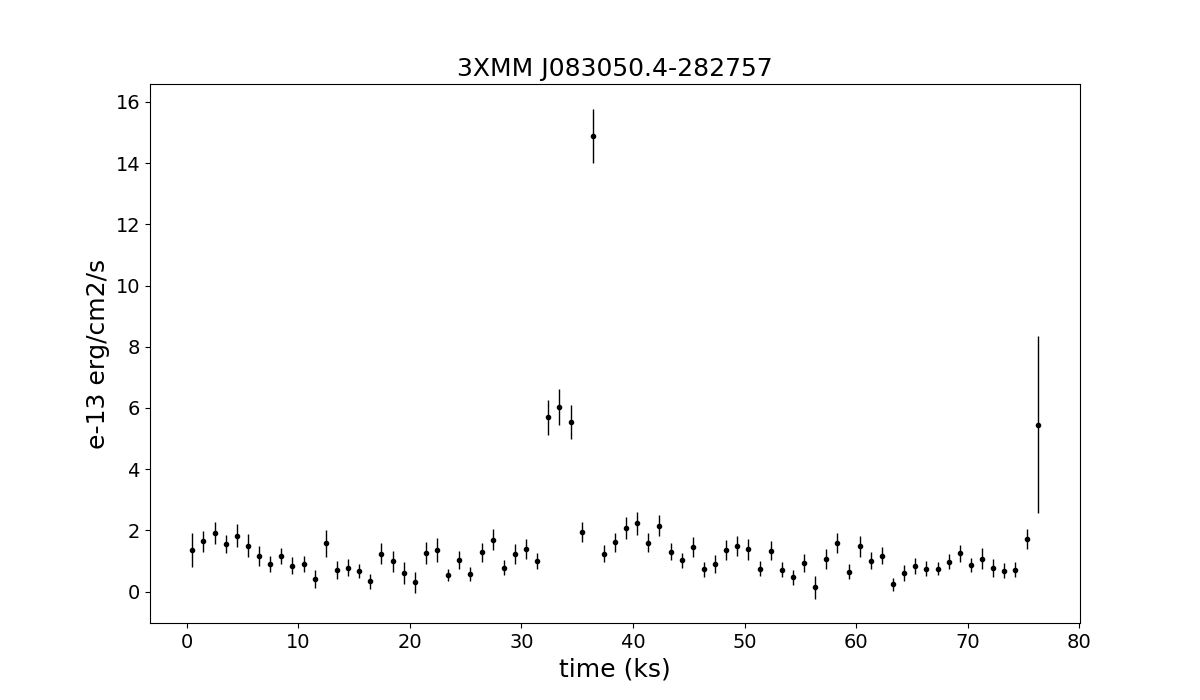}
    \caption{X-ray light curve of 3XMM\, J083850.4$-$282757, with a binning of size 997 s, in which appears a bright flare of duration of about 600s, see also \citet{Halpern2017a}.}
    \label{fig:0838_X}
\end{figure}

Using an updated version of the {\em EXTraS} tools \citep{Marelli2017}, we also searched for both flares and other types of short-term aperiodic variability in X-ray sources which are always in an "on" state within the same observation. Briefly, these tools employ a quantitative analysis of the X-ray light curve to spot deviations from a constant flux level as well as possible trends, such as flux modulations caused by either an intrinsic source periodicity or by the source eclipse. 
Only X-ray sources with at least 100 counts in the combination of the PN/MOS1/MOS2 detectors and classified as point-like according to their 3XMM/DR8 extension parameter are considered in the  {\em EXTraS} analysis. 
We found a bright X-ray flare only for 3XMM\, J083850.4$-$282757, associated with the RB candidate 4FGL\, J0838.7$-$2827, with a flare duration of $\sim$ 600s (see Fig. \ref{fig:0838_X}). This is the same flaring source pinpointed by \citet{Halpern2017a}, which proves that the automatic {\em EXTraS} procedure that we applied in the search for flares works effectively.

\subsubsection{Periodic X-ray variability}
\begin{figure}
    \centering
    \includegraphics[width=0.52\textwidth]{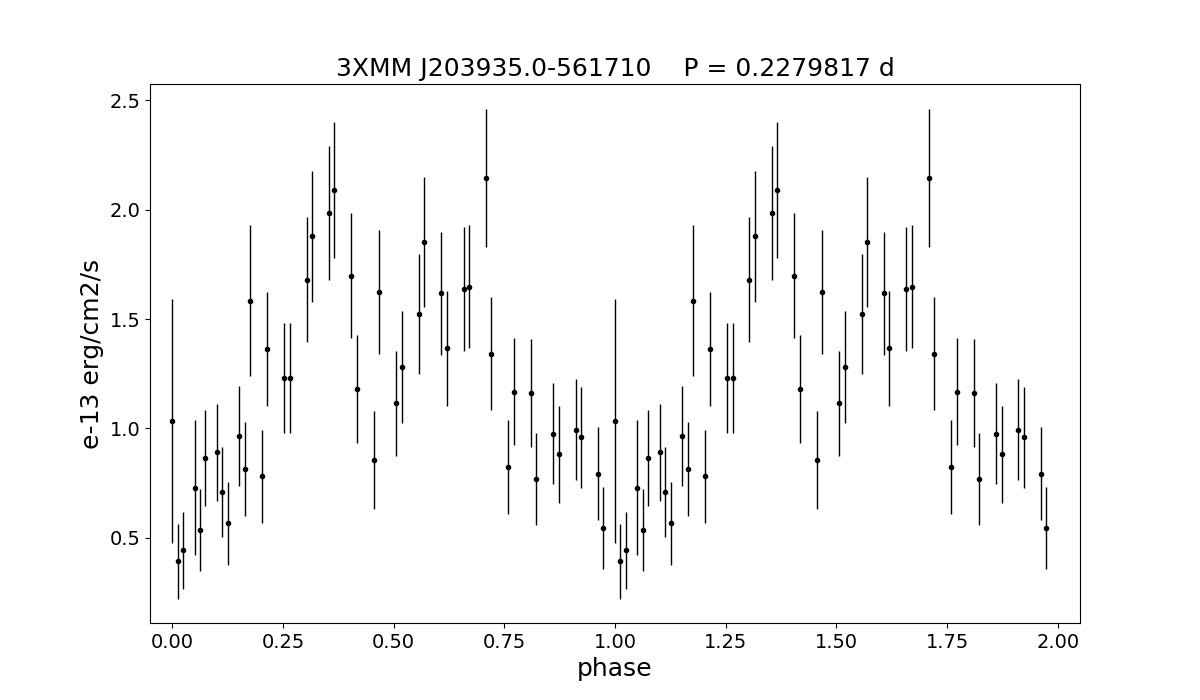}
    \caption{X-ray light curve of 3XMM\, J203935.0$-$561710, the X-ray counterpart to the $\gamma$-ray source 4FGL\, J2039.5$-$5617 folded at the orbital period computed by \citet{Strader2019}.}
    \label{fig:2039_X}
\end{figure}

\begin{figure}
\centering
\includegraphics[width=0.52\textwidth]{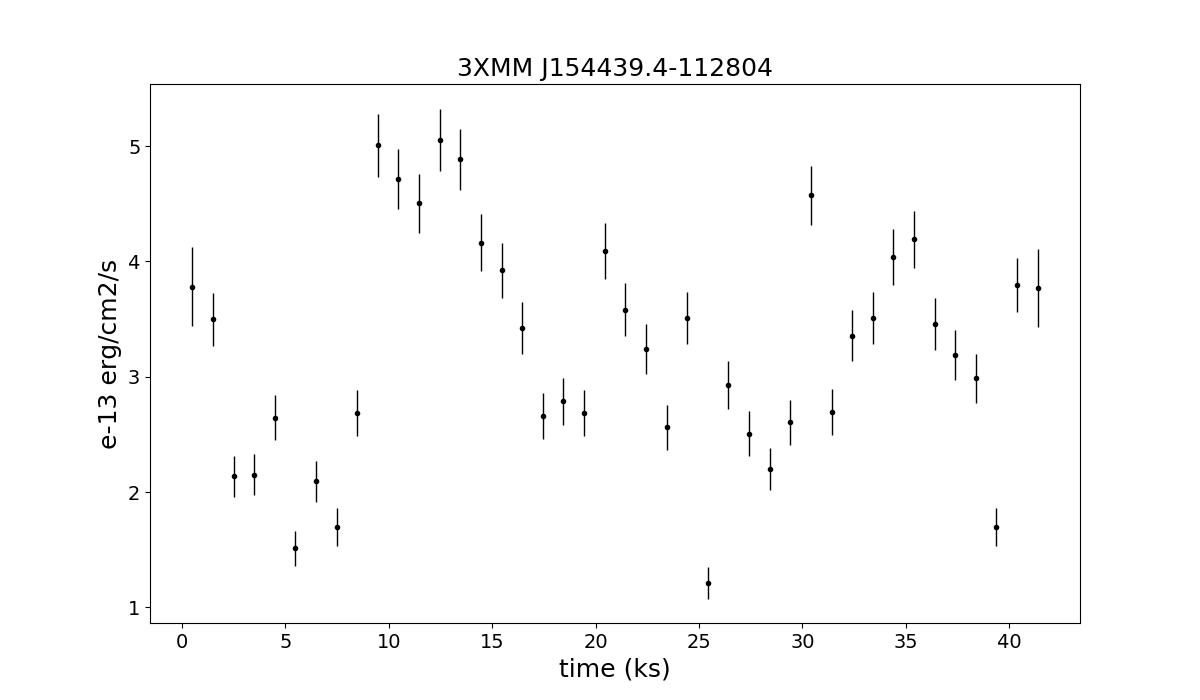} 
\includegraphics[width=0.52\textwidth]{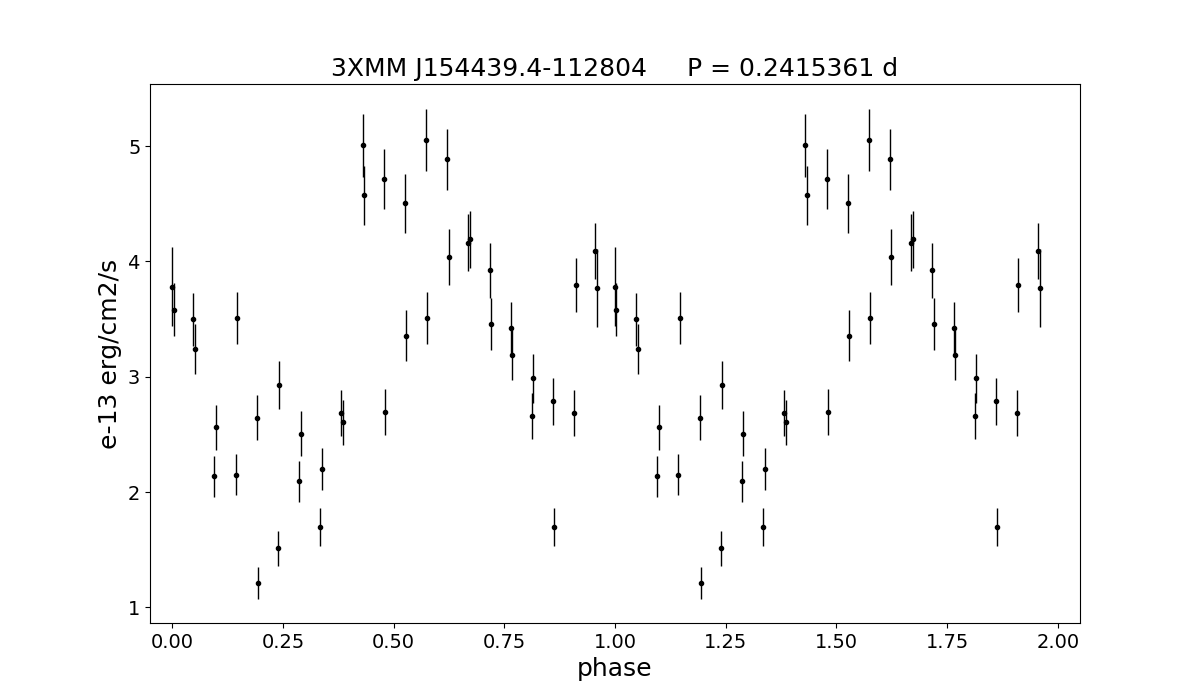} 
\caption{(top) Unfolded X-ray light curve of the X-ray source 3XMM\, J154439.4$-$112804, with a binning of size 997 s.(bottom) X-ray light curve folded at the orbital period of the binary system
\citep{Britt2017}. }
\label{fig:1544_X}
\end{figure}

We used the {\em EXTras} tools also to search for {\em XMM-Newton} sources with evidence of periodic variability, starting from the counterparts to the  known BW/RB candidates (Table \ref{tab:my-table}). Only four of them have been detected by {\em XMM-Newton}.
In the same 4FGL\, J0838.7$-$2827 field, we recovered the orbital and super-orbital flux modulations observed in the X-ray source 3XMM\, J083843.3$-$282701 (OBS ID 0764420101 and 0790180101), identified with the Cataclysmic Variable Star 1RXS\, J083842.1$-$282723 
\citep{Halpern2017a}. However, we did not find evidence of periodic flux modulations in 3XMM\, J083850.4$-$282757, the one associated with the $\gamma$-ray source, before and after subtracting the contribution of the bright X-ray flare.
As for the known BW/RB candidates with a measured optical periodicity (Table \ref{tab:periodi}), we recovered the periodic X-ray flux modulation ($\sim 0.22$d) in 3XMM\, J203935.0$-$561710 (OBS ID 0720750301), the X-ray counterpart to the RB candidate 4FGL\, J2039.5$-$5617 \citep{Salvetti2017}. The modulation (see Fig. \ref{fig:2039_X}) is more clearly recognised after folding the X-ray data at the orbital period of the binary system measured by \citet{Strader2019}.
We also found a trend of a more or less regular X-ray flux modulation in the unfolded light curve of 3XMM\, J154439.4$-$112804 (OBS ID 0724080101; 42.2 ks), as shown in Fig. \ref{fig:1544_X} (top), the X-ray counterpart to the candidate tMSP 4FGL\, J1544.5$-$1126 \citep{Bogdanov2015}. The 
X-ray light curve folded at 
the orbital period of the binary system, P$_{\rm orb}$=0.2415361 d
\citep{Britt2017} is also shown in Fig. \ref{fig:1544_X} (bottom).  Most likely, however
this modulation is due to 
the X-ray flux bi-modality observed during the observation \citep{Bogdanov2015}, when the source was in a low-luminosity accretion state.  
Observations obtained in a non-accreting state would give a better chance to discover a genuine orbital periodicity in the X-ray flux of 3XMM\, J154439.4$-$112804.
Finally, for the binary MSP and possible BW candidate 4FGL\, J1946.5$-$5402 (PSR\, J1946$-$5403) we reported a tentative optical flux modulation in its possible counterpart SSS\, J194633.7$-$540236 but only after folding the Catalina data at the orbital period of the binary system (0.13 d; Sectn.\, \ref{subsec:oldcand}). We failed to find clear evidence of flux periodicity in the associated X-ray source 3XMM\, J194633.6$-$540236 (OBS ID 0784771001;  $\sim$20 ks), with only a possible hint of a  $\sim$ 0.13 d modulation recognised in the unfolded light curve. An {\em XMM-Newton} observation longer than the one currently available is needed to fully cover at least two orbital cycles and investigate this possible periodicity.

We searched for evidence of X-ray periodicity among the six candidate $\gamma$-ray MSPs with a possible {\em XMM-Newton} counterpart
For 3XMM\, J210350.0$-$111338 (OBSID 0041150101 and 0041150201)
we only see a very marginal evidence of variability in the X-ray light curve.
For the remaining five candidate $\gamma$-ray MSPs with a possible {\em XMM-Newton} counterpart, with or without an associated optical counterpart (4FGL\, J0802.1$-$5612, 4FGL\, J1120.0$-$2204, 4FGL\, J1625.1$-$0020, 4FGL\, J2112.5$-$3043, and 4FGL\, J2333.1$-$5527), we found no evidence of periodicity in our blind search. The first four of these X-ray sources were preliminary searched for variability/periodicity by \citet{Salvetti2017} who found no evidence in either directions. Our analysis based on the {\em EXTraS} tools confirms their results.

Of course, we did not find any evidence of orbital periodicity in the {\em XMM-Newton} data of the three known isolated $\gamma$-ray pulsars (Table\, \ref{tab:my-table}), i.e. the young pulsar PSR\, J0359+5414 (4FGL\, J0359.4+5414) and the two MSPs PSR\, J1035$-$6720 (4FGL\, J1035.4$-$6720) and PSR\, J1744$-$76194 (4FGL\, J1744.0$-$7618), which we used as a yardstick to sort out doubtful cases. 

\subsubsection{Long-term X-ray variability}

Finally, we searched for possible long-term X-ray variability for all the 41 {\em XMM-Newton} sources in our sample.
Since it was not possible for us to re-run off-line the dedicated {\em EXTraS} long-term X-ray variability analysis, we browsed the dedicated on-line {\em EXTras} data products archive, which is still based on observations included in the 3XMM/DR5.  For this reason, for some sources we could not find corresponding data products. For the others we found that none deviates from a steady flux in any of the analysed energy bands. The lack of evidence of long-term X-ray variability  can be explained in some cases by the steady source nature, in others by an insufficiently long multi-epoch coverage.

\section{Summary and Conclusions}
\label{subsec:summ}

Using archival data and source catalogues, we carried out a multi-wavelength survey of the 48 MSP candidates selected  from unassociated 3FGL sources by \citet{SazParkinson2016}, based on machine-learning techniques. We found that 23 of these MSP-like $\gamma$-ray sources have candidate X-ray counterparts in {\em XMM-Newton}, {\em Chandra}, or {\em Swift}\footnote{When  our manuscript was about to be submitted an updated version of the {\em Swift} X-ray source catalogue, \citep[2SXPS;][] {Evans2020}, was published. We will update our work in the future using this catalogue as well as new releases of the {\em XMM-Newton} and {\em Chandra} X-ray source catalogues.}, of which 17 are associated with an optical counterpart in at least one of the multi-epoch surveys. 
Six of them show evidence of optical periodicity with a period smaller than 1 d, detected through a blind search.  We could recover the known optical periodicity for four confirmed BWs/RBs (4FGL\, J0523.3$-$2527, 4FGL\, J0955.3$-$3949, 4FGL\, J1653.6$-$0158, 4FGL\, J2039.5$-$5617), which proves the validity of our analysis. For  
two MSP candidates (4FGL\, J1627.7+3219
and 4FGL\, J2212.4+0708), we found a candidate optical periodicity for the first time, which needs to be confirmed by follow-up observations, 
making them BW/RB candidates. \\
We also revisited the optical identifications for two binary MSP candidates of \citet{SazParkinson2016}, now 4FGL\, J0802.1$-$5612 and 4FGL\, J0744.0$-$2525, proposed in \citet{Salvetti2017} on the basis of a periodic flux modulation. For the former, the proposed candidate counterpart now falls far from the updated $\gamma$-ray error ellipse and cannot be associated with 4FGL\, J0802.1$-$5612 any longer. For the latter, the candidate counterpart only falls marginally outside ($\sim 15\arcsec$) the 4FGL error circle and the association cannot be firmly ruled out. The periodicity could not be detected in the sparse multi-epoch optical survey data, so that we cannot investigate any long-term evolution of the orbit. An X-ray detection would be crucial to determine the nature of the proposed optical counterpart and verify its association with 4FGL\, J0744.0$-$2525. 

We made use of the {\em EXTraS} tools to run an X-ray variability analysis of our $\gamma$-ray MSP candidates in the {\em XMM-Newton} data, including the search for orbital periodicity and flares. Although we could recover known phenomena, e.g. the X-ray orbital modulation in  the RB candidate 4FGL\, J2039.5$-$5617 and the X-ray flare in the RB candidate 4FGL\, J0838.7$-$2827, these remain the only clear cases. Not surprisingly, the detection of X-ray modulations with a few hour period requires comparably long observations, whereas the detection of X-ray flares would benefit of a regular  monitoring.

Our BW/RB identification score (four confirmed plus 
two candidates out of the 23 examined MSP-like $\gamma$-ray sources)
would be higher than expected from the fraction of binary MSPs that are firmly identified as BWs/RBs  ($\approx$20\%) even if one assumes that the $\gamma$-ray sources in our sample are all MSPs, with at least one known exception
\citep[i.e., 4FGL\, J0359.4+5414
identified with the isolated young pulsars PSR\, J0359+5414;][]{Clark2017}, 
and that the corresponding fraction of binary ones is the same as in the entire MSP population ($\approx$65\%). Our identification score would be even more unexpected if one considers that our identification efficiency is affected by observational biases, such as the multi-wavelength coverage extent (both spatially and temporally)  of the $\gamma$-ray error ellipses of the MSP candidates (Sectn.\ \ref{subsec:xcorr}).  Interestingly, in  $\gamma$-rays the number of  identified BWs/RBs relative to the total number of binary MSPs is $\approx 40\%$, i.e. above the overall $\approx$20\% fraction, implying that a sample selection based on the $\gamma$-ray detection may introduce a statistical bias.  
As previously said, however, the identification of our 
two new BW/RB candidates still awaits confirmation, which might downplay the statistical impact of our results.

We plan to update our work once new MSP candidates are selected by machine-learning techniques \citep[e.g.,][]{SazParkinson2016} from new {\em Fermi} source catalogue releases, starting from the 4FGL,   exploiting the growing  multi-wavelength survey databases and a more systematic mining of archival data.
One of the reference for future works would be the list of $\gamma$-ray MSP candidates from the 4FGL published by \citet{Luo2020} when this works had been just finalised.
On a longer run, future systematic searches will greatly benefit from the data stream of X-ray/optical survey facilities, such as {\em e-Rosita} on the {\em Spectrum R\"ontgen Gamma} satellite, launched in July 2019,  and the Large Synoptic Survey Telescope, now Vera C. Rubin Observatory, with routine observations scheduled start in 2022.

\onecolumn
\begin{landscape}
\setlength\LTcapwidth{\linewidth}
\begin{longtable}{c|c|ccc|cccc}
\caption{Summary of the results of the multi-wavelength  cross-correlations of the 48 $\gamma$-ray MSP candidates selected in Sectn.\, \ref{subsec:strat}. Those with an least a candidate X-ray counterpart are 23, of which 17 have an associated optical counterpart. Columns identify the 4FGL name of the $\gamma$-ray source, the source identification, for the ten sources for which it is available, as isolated young pulsar (iPSR), isolated MSP (iMSP), candidate transitional MSP (ctMSP), binary MSP (bMSP), candidate BW or RB (cBW/cRB), the candidate X-ray counterparts in the selected X-ray catalogues (3XMM, 1SXPS, 2CXO) and their associated optical counterparts in the multi-epoch surveys (Catalina, PTF, ZTF, PanSTARRS), if any. In the PanSTARRS catalogue the same object can appear with multiple IDs corresponding to its detection in different observations. For the sake of clarity and compactness we only show one ID for each object. Optical counterparts with a confirmed/candidate periodicity (P) are marked by the superscript. The MSP candidates studied in \citet{Salvetti2017} are marked by the asterisk.}
\label{tab:my-table}
\\\hline
4FGL  & Id. & 3XMM & 1SXPS & 2CXO & CSS & PTF & ZTF  & PanSTARRS  \\ 
\hline
\endfirsthead

\hline
4FGL & Id & 3XMM & 1SXPS & 2CXO & CSS  & PTF  & ZTF  & PanSTARRS  \\ 
\hline
\endhead

\hline
\endfoot

\endlastfoot

J0212.1+5321 & - & - & - & J021214.0+531920 & - & - & 776208300032329 & -\\
 &  &  &  &  &  &  & 775205400044594 &  \\
 &  &  &  & J021215.8+531924 & - & - & - & - \\
 \hline
J0359.4+5414 & iPSR & J035925.2+541455 & - & J035926.0+541455 & - & - & - & -\\
 &  & - & - & J035925.4+541501 & - & - & 778207100035977 & - \\
 &  &  &  & J035922.8+541601 & - & - & 778207100048011 & - \\
 &  & J035927.3+541358 & - & J035927.2+541356 & - & - & - & - \\
 &  & - & - & J035929.7+541555 & - & 48822060004599 & 778107100010205 & 173110598739469282 \\
 &  & - & - & J035926.0+541618 & - & - & - & -\\
 \hline
J0523.3$-$2527$^*$ & cRB & - & J052317.0$-$252732 & - & J052316.9$-$252737$^{\rm P}$ & - & 256107300002741 & 77440808205337977 \\
 &  & - & J052324.2$-$252732 & - & J052324.4$-$252731 & - & 256207300002270 & 77440808516949749 \\
 \hline
J0802.1$-$5612$^*$ & - & J080203.6$-$561352 & - & - & J080203.6$-$561354  & - & - & - \\
 &  & J080217.0$-$561444 & - & - & J080216.8$-$561444 & - & - & - \\
 &  & J080201.5$-$561142 & - & - & - & - & - & -\\
 &  & J080208.5$-$561216 & - & - & - & - & - & -\\
 &  & J080214.0$-$561325 & - & - & - & - & - & -\\
 &  & J080219.2$-$561438 & - & - & - & - & - & -\\
 \hline
J0838.7$-$2827 & cRB & J083842.9$-$282830 & J083842.4$-$282830 & - & - & - & - & 73821296782839968 \\
 &  & J083843.2$-$282546 & - & - & - & - & - & -\\
 &  & J083843.3$-$282701 & J083843.1$-$282702 & - & - & - & - & 73851296805669950\\
 &  & J083844.5$-$282831 & - & - & - & - & - & 73831296857810040 \\
 &  & J083846.9$-$282646 & - & - & - & - & - & 73861296944754027 \\
 &  & J083847.7$-$282702 & - & - & - & - & - & - \\
 &  & J083850.4$-$282757 & - & - & - & - & - & 73841297100561268 \\
 &  & J083847.5$-$282837 & - & - & - & - & - & - \\
 \hline
 J0933.8$-$6232 & - & - & - & J093344.5$-$623317 & - & - & - & - \\
 &  & - & - & J093400.6$-$623352  & - & - & - & - \\
 \hline
J0955.3$-$3949 & cRB & - & J095527.8$-$394750 & - & J095527.8$-$394752$^{\rm P}$ & - & - & - \\
 \hline
 J1035.4$-$6720$^*$ & iMSP & J103527.3$-$672013	& J103526.9$-$672010 & - & - & - & - & - \\
 \hline
J1120.0$-$2204$^*$ & - & J111958.3$-$220456 & J111958.2$-$220458 & - & J111958.3$-$220456 & - & 268115400002304 & 81501699930311522 \\
 &  & J112001.7$-$220457 & - & - & J112001.7$-$220457 & - & 268215400011285 & 81501700074071334 \\
 \hline
J1539.4$-$3323$^*$ & - & - & J153924.7$-$332233	& - & - & - & - & - \\
 \hline
J1544.5$-$1126 & ctMSP & J154425.0$-$112602 & - & - & - & - & - & - \\
 &  & J154431.4$-$112451 & - & - & - & 221372090010448 & - & 94302361306503089 \\
 &  & J154437.8$-$112659 & - &  & - & - & - & - \\
 &  & J154438.5$-$112634 & J154438.5$-$112641 & - & J154438.5-112635 & 221372090010841 & 378208400009125 & 94262361606388530 \\
 &  & J154439.4$-$112804 & J154439.3$-$112804 & - & J154439.4-112804 & - & 378108400011640 & 94232361640998794 \\
 &  & J154440.0$-$112741 & - & - & - & 221372090010994 & 378208400009349 & 94242361661476451 \\
 &  & J154441.0$-$112557 & - & - & - & - & - & - \\
 &  & J154441.1$-$112456 & - & - & - & - & 378208400008746 & 94302361717401469 \\

 \hline
J1625.1$-$0020$^*$ & - & J162509.4$-$002052 & - & - & J162509.5$-$002051 & - & - & -\\
 &  & J162510.3$-$002127 & - & - & - & - & - & - \\
 \hline
J1627.7+3219 & - & - & J162742.8+322059 & - & J162743.0+322103$^{\rm P}$ & 41621060004671$^{\rm P}$ & 679105100004944 & 146822469291741697 \\
 \hline
J1653.6$-$0158$^*$ & cBW & - & J165337.8$-$015835 & J165338.0$-$015836 & J165338.1$-$015836$^{\rm P}$ & 26652110004413 & 432112400014707 & 105622534086488129 \\
 &  & - & - & J165334.8$-$015803 & - & 26652115002981 & - & 105632533941618183 \\
 \hline
J1744.0$-$7618$^*$ & iMSP & J174400.6$-$761914 & - & - & - & - & - & - \\
 \hline
J1946.5$-$5402 & bMSP$\dagger$ & J194633.6$-$540236 & - & - & J194633.7$-$540236 & - & - & - \\
 &  & J194634.4$-$540343 & - & - & - & - & - & - \\
 \hline
J2004.3+3339 & - & - & - & J200421.3+333942 & - & - & 686209300101711 & 148393010879214642 \\
 &  &  &  &  &  &  & 686209300084182 & 148393010887423147 \\
 &  & - & - & J200422.7+333844 & - & - & 686109300049900 & 148373010937876456 \\
  &  &  &  &  &  &  & 686209300122014 & 148373010954324969\\
 &  & - & - & J200422.9+333843 & - & - & 686209300122014 & 148373010954324969 \\
 &  & - & - & J200423.4+333906 & - & - &  686209300131114 & 148383010978953660 \\
 &  &  &  &  &  &  &  686109300072405 & 148383010985852102 \\
  &  &  &  &  &  &  &  686109300089082 & 148383010967343943 \\
 &  & - & - & J200424.0+334008 & - & - & 686209300032950 & 148403011000363517 \\
 &  & - & - & J200427.9+333857 & - & - & 686209300128511 & 148373011155519463 \\
 &  &  &  &  &  &  & 686209300033911 & 148383011165801287 \\
 \hline
J2039.5$-$5617$^*$ & cRB & J203935.0$-$561710 & - & - & J203934.9$-$561708$^{\rm P}$ & - & - & - \\
 \hline
J2103.7$-$1112 & - & J210344.8$-$111153 & - & - & - & - & - & 94563159378063313 \\
 &  & J210350.0$-$111338 & - & J210350.0$-$111341 & J210350.1$-$111341
  & 22642070000705 & 389106300020270 & 94523159586916709 \\
 \hline
J2112.5$-$3043$^*$ & - & J211232.1$-$304403	& - & - & - & - & - \\
 \hline
J2133.1$-$6432 & - & - & J213314.9$-$643229 & - & - & - & - & - \\
 \hline
J2212.4+0708 & - & - &  J221230.8+070651 & - & J221231.0+070652$^{\rm P}$ & 31092070002977 & 494213400002994 & 116533331290607966 \\
 \hline
J2333.1$-$5527 & - & J233300.1$-$552549	& - & -	& -	& -	& -	& - \\
 &  & J233304.1$-$552854 & - & - & - & - & - & - \\
 &  & J233313.8$-$552814 & - & - & - & - & - & - \\
 &  & J233316.0$-$552620 & - & - & J233315.9$-$552620 & - & - & - \\
 &  & J233317.3$-$552554 & - & - & - & - & - & - \\
 \hline
\end{longtable}
\begin{tablenotes}
\small
\item{$\dagger$ 4FGL\, J1946.5-5402 is identified with a binary MSP (PSR\, J1946$-$5403) which is is considered a possible BW candidate \citep{Camilo2015}.}
\end{tablenotes}
\end{landscape}

\begin{landscape}
\begin{table}
\centering
\caption{Summary of the confirmed/candidate $\gamma$-ray MSPs listed in Table \ref{tab:my-table} with either known or candidate orbital periodicity. X-ray counterparts are listed in columns 3--5. Sources for which the companion star has been optically identified from previous works are marked by the superscript (O). For all of them, the value of the orbital period is known (superscript PB) and has been measured in the optical, either from photometry or spectroscopy measurements.  
The values of the orbital period computed in this work from the LS periodogram analysis of the optical light curves from the survey data (columns 6 and 7) are listed in the column 9 together with the associated uncertainties. The computed significance of the corresponding periodogram peak is listed in the last column. Periods in agreement with values reported in the literature are marked in bold and candidate periods in italics. Orbital periods which we could not recover through our LS periodogram analysis are marked in roman. For 4FGL\, J0838.7$-$2827 and 4FGL\, J1544.5$-$1126 the orbital period has been measured from optical photometry and spectroscopy, respectively, whereas for 4FGL\, J1946.5$-$5402 (PSR\, J1946$-$5403)  from radio timing observations. }
\label{tab:periodi}
\begin{threeparttable}
\begin{tabular}{lccccccccccc}
\hline
4FGL & Id. & 3XMM & 1SXPS & 2CXO & Catalina & PTF & $<$V$>$  & Period (d) & Significance \\
\hline
J0523.3$-$2527$^{\rm O,PB}$ & cRB\tnote{1} & - & J052317.0$-$252732 & - & J052316.9$-$252737 &  & 16.51 & {\bf 0.68813(6)} & 7 $\sigma$\\
J0838.7$-$2827$^{\rm O,PB}$ & cRB\tnote{2} & J083850.4$-$282757 & - & - & - & - &  & 0.214507 & \\
J0955.3$-$3949$^{\rm O,PB}$ & cRB\tnote{3} & - & J095527.8$-$394750 & - & J095527.8$-$394752 &  & 18.45 & {\bf 0.38733(6)} & 7 $\sigma$\\
J1544.5$-$1126$^{\rm O,PB}$ & ctMSP\tnote{4} & J154439.4$-$112804 & J154439.3$-$112804 & - & J154439.4$-$112804 &  & 18.18 & 0.2415361 & \\
J1627.7+3219 & - & - &J162742.8+322059 & - & J162743.0+322103 & 41621060004671 & 13.35 & {\it 0.49927(2)} & 7 $\sigma$ \\
J1653.6$-$0158$^{\rm O,PB}$ & BW\tnote{5}& - & J165337.8$-$015835 & J165338.0$-$015836 & J165338.1$-$015836 & & 20.26 & {\bf 0.054799(2)} & 7 $\sigma$\\
J1946.5$-$5402 & bMSP\tnote{6}& J194633.6$-$540236 & - & - & J194633.7$-$540236 & & 19.31 & 0.13 & \\
J2039.5$-$5617$^{\rm O,PB}$ & cRB\tnote{7} & J203935.0$-$561710 & - & - & J203934.9$-$561708 &  & 18.54 & {\bf 0.22798(3)} & 7 $\sigma$\\
J2212.4+0708   & - &        -         & J221230.8+070651 & - & J221231.0+070652 &  & 19.70      &{\it 0.31884(6) } & 3.2 $\sigma$        \\
\hline
\end{tabular}
\begin{tablenotes}
\item{\bf References:} (1) \citet{Strader2014}; (2) \citet{Halpern2017b}; (3) \citet{Li2018}; (4) \citet{Britt2017}; (5) \citet{Romani2014}; (6) \citet{Camilo2015}; (7) \citet{Salvetti2015}
\end{tablenotes}
\end{threeparttable}
\end{table}
\end{landscape}
\newpage
\twocolumn

\section*{Acknowledgements}
\addcontentsline{toc}{section}{Acknowledgements}
We thank the anonymous referee for her/his positive and constructive comments to the manuscript.
CB acknowledges hospitality from INAF-IASF, Milan during the period of her MSc Thesis work. 
This research has made use of data obtained from the 3XMM {\em XMM-Newton} serendipitous source catalogue compiled by the 10 institutes of the {\em XMM-Newton} Survey Science Centre selected by ESA.
This research has made use of data obtained from the Chandra Source Catalog, provided by the {\em Chandra} X-ray Center (CXC) as part of the {\em Chandra} Data Archive.  
The CSS survey is funded by the National Aeronautics and Space
Administration under Grant No. NNG05GF22G issued through the Science
Mission Directorate Near-Earth Objects Observations Program.  The CRTS
survey is supported by the U.S.~National Science Foundation under
grants AST-0909182.
The Pan-STARRS1 Surveys (PS1) and the PS1 public science archive have been made possible through contributions by the Institute for Astronomy, the University of Hawaii, the Pan-STARRS Project Office, the Max-Planck Society and its participating institutes, the Max Planck Institute for Astronomy, Heidelberg and the Max Planck Institute for Extraterrestrial Physics, Garching, The Johns Hopkins University, Durham University, the University of Edinburgh, the Queen's University Belfast, the Harvard-Smithsonian Center for Astrophysics, the Las Cumbres Observatory Global Telescope Network Incorporated, the National Central University of Taiwan, the Space Telescope Science Institute, the National Aeronautics and Space Administration under Grant No. NNX08AR22G issued through the Planetary Science Division of the NASA Science Mission Directorate, the National Science Foundation Grant No. AST-1238877, the University of Maryland, Eotvos Lorand University (ELTE), the Los Alamos National Laboratory, and the Gordon and Betty Moore Foundation.
This research has made use of data produced by the EXTraS project, funded by the European Union's Seventh Framework Programme under grant agreement no 607452

%%%%%%%%%%%%%%%%%%%%%%%%%%%%%%%%%%%%%%%%%%%%%%%%%%



\bibliographystyle{mnras}
\bibliography{bibliography} % if your bibtex file is called example.bib

\begin{thebibliography}{}
\makeatletter
\relax
\def\mn@urlcharsother{\let\do\@makeother \do\$\do\&\do\#\do\^\do\_\do\%\do\~}
\def\mn@doi{\begingroup\mn@urlcharsother \@ifnextchar [ {\mn@doi@}
  {\mn@doi@[]}}
\def\mn@doi@[#1]#2{\def\@tempa{#1}\ifx\@tempa\@empty \href
  {http://dx.doi.org/#2} {doi:#2}\else \href {http://dx.doi.org/#2} {#1}\fi
  \endgroup}
\def\mn@eprint#1#2{\mn@eprint@#1:#2::\@nil}
\def\mn@eprint@arXiv#1{\href {http://arxiv.org/abs/#1} {{\tt arXiv:#1}}}
\def\mn@eprint@dblp#1{\href {http://dblp.uni-trier.de/rec/bibtex/#1.xml}
  {dblp:#1}}
\def\mn@eprint@#1:#2:#3:#4\@nil{\def\@tempa {#1}\def\@tempb {#2}\def\@tempc
  {#3}\ifx \@tempc \@empty \let \@tempc \@tempb \let \@tempb \@tempa \fi \ifx
  \@tempb \@empty \def\@tempb {arXiv}\fi \@ifundefined
  {mn@eprint@\@tempb}{\@tempb:\@tempc}{\expandafter \expandafter \csname
  mn@eprint@\@tempb\endcsname \expandafter{\@tempc}}}

\bibitem[\protect\citeauthoryear{{Abdollahi} et~al.,}{{Abdollahi}
  et~al.}{2020}]{4fgl}
{Abdollahi} S.,  et~al., 2020, \mn@doi [\apjs] {10.3847/1538-4365/ab6bcb},
  \href {https://ui.adsabs.harvard.edu/abs/2020ApJS..247...33A} {247, 33}

\bibitem[\protect\citeauthoryear{Acero et~al.,}{Acero et~al.}{2015}]{Acero2015}
Acero F.,  et~al., 2015, \mn@doi [The Astrophysical Journal Supplement Series]
  {10.1088/0067-0049/218/2/23}, 218, 23

\bibitem[\protect\citeauthoryear{Alpar, Cheng, Ruderman  \& Shaham}{Alpar
  et~al.}{1982}]{Alpar1982}
Alpar M.~A.,  Cheng A.~F.,  Ruderman M.~A.,   Shaham J.,  1982, \mn@doi
  [Nature] {10.1038/300728a0}, 300, 728

\bibitem[\protect\citeauthoryear{Archibald et~al.,}{Archibald
  et~al.}{2009}]{Archibald2009}
Archibald A.~M.,  et~al., 2009, \mn@doi [Science] {10.1126/science.1172740},
  324, 1411

\bibitem[\protect\citeauthoryear{{Atwood} et~al.,}{{Atwood}
  et~al.}{2009}]{Atwood2009}
{Atwood} W.~B.,  et~al., 2009, \mn@doi [\apj] {10.1088/0004-637X/697/2/1071},
  \href {https://ui.adsabs.harvard.edu/abs/2009ApJ...697.1071A} {697, 1071}

\bibitem[\protect\citeauthoryear{Backer, Kulkarni, Heiles, Davis  \&
  Goss}{Backer et~al.}{1982}]{Backer1982}
Backer D.~C.,  Kulkarni S.~R.,  Heiles C.,  Davis M.~M.,   Goss W.~M.,  1982,
  \mn@doi [Nature] {10.1038/300615a0}, 300, 615

\bibitem[\protect\citeauthoryear{Bellm et~al.,}{Bellm et~al.}{2018}]{Bellm2018}
Bellm E.~C.,  et~al., 2018, \mn@doi [Publications of the Astronomical Society
  of the Pacific] {10.1088/1538-3873/aaecbe}, 131, 018002

\bibitem[\protect\citeauthoryear{Bogdanov \& Halpern}{Bogdanov \&
  Halpern}{2015}]{Bogdanov2015}
Bogdanov S.,  Halpern J.~P.,  2015, \mn@doi [The Astrophysical Journal]
  {10.1088/2041-8205/803/2/l27}, 803, L27

\bibitem[\protect\citeauthoryear{Bonanos et~al.,}{Bonanos
  et~al.}{2019}]{Bonanos2019}
Bonanos A.~Z.,  et~al., 2019, \mn@doi [Astronomy {\&} Astrophysics]
  {10.1051/0004-6361/201936026}, 630, A92

\bibitem[\protect\citeauthoryear{Britt, Strader, Chomiuk, Tremou, Peacock,
  Halpern  \& Salinas}{Britt et~al.}{2017}]{Britt2017}
Britt C.~T.,  Strader J.,  Chomiuk L.,  Tremou E.,  Peacock M.,  Halpern J.,
  Salinas R.,  2017, \mn@doi [The Astrophysical Journal]
  {10.3847/1538-4357/aa8e41}, 849, 21

\bibitem[\protect\citeauthoryear{{Bruel}, {Burnett}, {Digel}, {Johannesson},
  {Omodei}  \& {Wood}}{{Bruel} et~al.}{2018}]{Bruel2018}
{Bruel} P.,  {Burnett} T.~H.,  {Digel} S.~W.,  {Johannesson} G.,  {Omodei} N.,
   {Wood} M.,  2018, arXiv e-prints, \href
  {https://ui.adsabs.harvard.edu/abs/2018arXiv181011394B} {p. arXiv:1810.11394}

\bibitem[\protect\citeauthoryear{Camilo et~al.,}{Camilo
  et~al.}{2015}]{Camilo2015}
Camilo F.,  et~al., 2015, \mn@doi [The Astrophysical Journal]
  {10.1088/0004-637x/810/2/85}, 810, 85

\bibitem[\protect\citeauthoryear{{Chambers} \& {Pan-STARRS Team}}{{Chambers} \&
  {Pan-STARRS Team}}{2016}]{Chambers2016}
{Chambers} K.~C.,  {Pan-STARRS Team} 2016, in American Astronomical Society
  Meeting Abstracts \#227. p. 324.07

\bibitem[\protect\citeauthoryear{{Cho}, {Halpern}  \& {Bogdanov}}{{Cho}
  et~al.}{2018}]{Cho2018}
{Cho} P.~B.,  {Halpern} J.~P.,   {Bogdanov} S.,  2018, \mn@doi [\apj]
  {10.3847/1538-4357/aade92}, \href
  {https://ui.adsabs.harvard.edu/abs/2018ApJ...866...71C} {866, 71}

\bibitem[\protect\citeauthoryear{Clark et~al.,}{Clark et~al.}{2017}]{Clark2017}
Clark C.~J.,  et~al., 2017, \mn@doi [The Astrophysical Journal]
  {10.3847/1538-4357/834/2/106}, 834, 106

\bibitem[\protect\citeauthoryear{Clark et~al.,}{Clark et~al.}{2018}]{Clark2018}
Clark C.~J.,  et~al., 2018, \mn@doi [Science Advances]
  {10.1126/sciadv.aao7228}, 4, eaao7228

\bibitem[\protect\citeauthoryear{Dai, Wang, Vadakkumthani  \& Xing}{Dai
  et~al.}{2016}]{Dai2016}
Dai X.-J.,  Wang Z.-X.,  Vadakkumthani J.,   Xing Y.,  2016, \mn@doi [Research
  in Astronomy and Astrophysics] {10.1088/1674-4527/16/6/097}, 16, 012

\bibitem[\protect\citeauthoryear{Dai, Wang, Vadakkumthani  \& Xing}{Dai
  et~al.}{2017}]{Dai2017}
Dai X.-J.,  Wang Z.-X.,  Vadakkumthani J.,   Xing Y.,  2017, \mn@doi [Research
  in Astronomy and Astrophysics] {10.1088/1674-4527/17/7/72}, 17, 072

\bibitem[\protect\citeauthoryear{{De Luca}, Salvaterra, Tiengo, D'Agostino,
  Watson, Haberl  \& Wilms}{{De Luca} et~al.}{2016}]{DeLuca2016}
{De Luca} A.,  Salvaterra R.,  Tiengo A.,  D'Agostino D.,  Watson M.~G.,
  Haberl F.,   Wilms J.,  2016, in , Astrophysics and Space Science
  Proceedings.
Springer International Publishing, pp 291--295,
  \mn@doi{10.1007/978-3-319-19330-4_46}, \url
  {https://doi.org/10.1007/978-3-319-19330-4_46}

\bibitem[\protect\citeauthoryear{Evans et~al.,}{Evans et~al.}{2010}]{Evans2010}
Evans I.~N.,  et~al., 2010, \mn@doi [The Astrophysical Journal Supplement
  Series] {10.1088/0067-0049/189/1/37}, 189, 37

\bibitem[\protect\citeauthoryear{Evans et~al.,}{Evans et~al.}{2013}]{Evans2013}
Evans P.~A.,  et~al., 2013, \mn@doi [The Astrophysical Journal Supplement
  Series] {10.1088/0067-0049/210/1/8}, 210, 8

\bibitem[\protect\citeauthoryear{{Evans} et~al.,}{{Evans}
  et~al.}{2020}]{Evans2020}
{Evans} P.~A.,  et~al., 2020, \mn@doi [\apjs] {10.3847/1538-4365/ab7db9}, \href
  {https://ui.adsabs.harvard.edu/abs/2020ApJS..247...54E} {247, 54}

\bibitem[\protect\citeauthoryear{Fruchter, Stinebring  \& Taylor}{Fruchter
  et~al.}{1988}]{Fruchter1988}
Fruchter A.~S.,  Stinebring D.~R.,   Taylor J.~H.,  1988, \mn@doi [Nature]
  {10.1038/333237a0}, 333, 237

\bibitem[\protect\citeauthoryear{Halpern, Bogdanov  \& Thorstensen}{Halpern
  et~al.}{2017a}]{Halpern2017a}
Halpern J.~P.,  Bogdanov S.,   Thorstensen J.~R.,  2017a, \mn@doi [The
  Astrophysical Journal] {10.3847/1538-4357/838/2/124}, 838, 124

\bibitem[\protect\citeauthoryear{Halpern, Strader  \& Li}{Halpern
  et~al.}{2017b}]{Halpern2017b}
Halpern J.~P.,  Strader J.,   Li M.,  2017b, \mn@doi [The Astrophysical
  Journal] {10.3847/1538-4357/aa7cff}, 844, 150

\bibitem[\protect\citeauthoryear{Hui \& Li}{Hui \& Li}{2019}]{Hui2019}
Hui C.~Y.,  Li K.~L.,  2019, \mn@doi [Galaxies] {10.3390/galaxies7040093}, 7,
  93

\bibitem[\protect\citeauthoryear{{Jaodand}, {Hessels}  \&
  {Archibald}}{{Jaodand} et~al.}{2018}]{Jaodand2018}
{Jaodand} A.,  {Hessels} J. W.~T.,   {Archibald} A.,  2018, in {Weltevrede} P.,
   {Perera} B.~B.~P.,  {Preston} L.~L.,   {Sanidas} S.,  eds,  IAU Symposium
  Vol. 337, Pulsar Astrophysics the Next Fifty Years. pp 47--51 (\mn@eprint
  {arXiv} {1711.10565}), \mn@doi{10.1017/S1743921317010407}

\bibitem[\protect\citeauthoryear{{Kulkarni}}{{Kulkarni}}{2013}]{Kulkarni2013}
{Kulkarni} S.~R.,  2013, The Astronomer's Telegram, \href
  {https://ui.adsabs.harvard.edu/abs/2013ATel.4807....1K} {4807, 1}

\bibitem[\protect\citeauthoryear{Lansbury et~al.,}{Lansbury
  et~al.}{2017}]{Lansbury2017}
Lansbury G.~B.,  et~al., 2017, \mn@doi [The Astrophysical Journal]
  {10.3847/1538-4357/aa8176}, 846, 20

\bibitem[\protect\citeauthoryear{{Larson}, {Beshore}, {Hill}, {Christensen},
  {McLean}, {Kolar}, {McNaught}  \& {Garradd}}{{Larson}
  et~al.}{2003}]{Larson2003}
{Larson} S.,  {Beshore} E.,  {Hill} R.,  {Christensen} E.,  {McLean} D.,
  {Kolar} S.,  {McNaught} R.,   {Garradd} G.,  2003, in AAS/Division for
  Planetary Sciences Meeting Abstracts \#35. AAS/Division for Planetary
  Sciences Meeting Abstracts.
p. 36.04

\bibitem[\protect\citeauthoryear{Li et~al.,}{Li et~al.}{2018}]{Li2018}
Li K.-L.,  et~al., 2018, \mn@doi [The Astrophysical Journal]
  {10.3847/1538-4357/aad243}, 863, 194

\bibitem[\protect\citeauthoryear{Linares}{Linares}{2014}]{Linares2014}
Linares M.,  2014, \mn@doi [The Astrophysical Journal]
  {10.1088/0004-637x/795/1/72}, 795, 72

\bibitem[\protect\citeauthoryear{Linares}{Linares}{2019}]{Linares:2019aua}
Linares M.,  2019, in {13th Frascati Workshop on Multifrequency Behaviour of
  High Energy Cosmic Sources (MULTIF2019) Palermo, Italy, June 3-8, 2019}.
  (\mn@eprint {arXiv} {1910.09572})

\bibitem[\protect\citeauthoryear{Luo, Leung, Hui  \& Li}{Luo
  et~al.}{2020}]{Luo2020}
Luo S.,  Leung A.~P.,  Hui C.~Y.,   Li K.~L.,  2020, \mn@doi [Monthly Notices
  of the Royal Astronomical Society] {10.1093/mnras/staa166}

\bibitem[\protect\citeauthoryear{{Marelli} et~al.,}{{Marelli}
  et~al.}{2017}]{Marelli2017}
{Marelli} M.,  et~al., 2017, \mn@doi [\apjl] {10.3847/2041-8213/aa9b2e}, \href
  {https://ui.adsabs.harvard.edu/abs/2017ApJ...851L..27M} {851, L27}

\bibitem[\protect\citeauthoryear{Mignani}{Mignani}{2011}]{Mignani2011}
Mignani R.~P.,  2011, \mn@doi [Advances in Space Research]
  {10.1016/j.asr.2009.12.011}, 47, 1281

\bibitem[\protect\citeauthoryear{Mignani et~al.,}{Mignani
  et~al.}{2014}]{Mignani2014}
Mignani R.~P.,  et~al., 2014, \mn@doi [Monthly Notices of the Royal
  Astronomical Society] {10.1093/mnras/stu1300}, 443, 2223

\bibitem[\protect\citeauthoryear{Ng, Takata, Strader, Li  \& Cheng}{Ng
  et~al.}{2018}]{Ng2018}
Ng C.~W.,  Takata J.,  Strader J.,  Li K.~L.,   Cheng K.~S.,  2018, \mn@doi
  [The Astrophysical Journal] {10.3847/1538-4357/aae308}, 867, 90

\bibitem[\protect\citeauthoryear{Pletsch et~al.,}{Pletsch
  et~al.}{2012}]{Pletsch2012}
Pletsch H.~J.,  et~al., 2012, \mn@doi [Science] {10.1126/science.1229054}, 338,
  1314

\bibitem[\protect\citeauthoryear{{Radhakrishnan} \&
  {Srinivasan}}{{Radhakrishnan} \& {Srinivasan}}{1982}]{Radha1982}
{Radhakrishnan} V.,  {Srinivasan} G.,  1982, Current Science, \href
  {https://ui.adsabs.harvard.edu/abs/1982CSci...51.1096R} {51, 1096}

\bibitem[\protect\citeauthoryear{Rau et~al.,}{Rau et~al.}{2009}]{Rau2009}
Rau A.,  et~al., 2009, \mn@doi [Publications of the Astronomical Society of the
  Pacific] {10.1086/605911}, 121, 1334

\bibitem[\protect\citeauthoryear{Ray et~al.,}{Ray et~al.}{2013}]{Ray2013}
Ray P.~S.,  et~al., 2013, \mn@doi [The Astrophysical Journal]
  {10.1088/2041-8205/763/1/l13}, 763, L13

\bibitem[\protect\citeauthoryear{{Ray} et~al.,}{{Ray} et~al.}{2016}]{Ray2016}
{Ray} P.~S.,  et~al., 2016, in American Astronomical Society Meeting Abstracts
  \#227. p. 423.07

\bibitem[\protect\citeauthoryear{{Ray} et~al.,}{{Ray} et~al.}{2020}]{Ray2020}
{Ray} P.~S.,  et~al., 2020, \mn@doi [Research Notes of the American
  Astronomical Society] {10.3847/2515-5172/ab7eb5}, \href
  {https://ui.adsabs.harvard.edu/abs/2020RNAAS...4...37R} {4, 37}

\bibitem[\protect\citeauthoryear{Rea et~al.,}{Rea et~al.}{2017}]{Rea2017}
Rea N.,  et~al., 2017, \mn@doi [Monthly Notices of the Royal Astronomical
  Society] {10.1093/mnras/stx1560}, 471, 2902

\bibitem[\protect\citeauthoryear{Roberts}{Roberts}{2012}]{Roberts2012}
Roberts M. S.~E.,  2012, \mn@doi [Proceedings of the International Astronomical
  Union] {10.1017/s174392131202337x}, 8, 127

\bibitem[\protect\citeauthoryear{Romani}{Romani}{2015}]{Romani2015}
Romani R.~W.,  2015, \mn@doi [The Astrophysical Journal]
  {10.1088/2041-8205/812/2/l24}, 812, L24

\bibitem[\protect\citeauthoryear{Romani, Filippenko  \& Cenko}{Romani
  et~al.}{2014}]{Romani2014}
Romani R.~W.,  Filippenko A.~V.,   Cenko S.~B.,  2014, \mn@doi [The
  Astrophysical Journal] {10.1088/2041-8205/793/1/l20}, 793, L20

\bibitem[\protect\citeauthoryear{Rosen et~al.,}{Rosen et~al.}{2016}]{Rosen2016}
Rosen S.~R.,  et~al., 2016, \mn@doi [Astronomy {\&} Astrophysics]
  {10.1051/0004-6361/201526416}, 590, A1

\bibitem[\protect\citeauthoryear{Salvetti et~al.,}{Salvetti
  et~al.}{2015}]{Salvetti2015}
Salvetti D.,  et~al., 2015, \mn@doi [The Astrophysical Journal]
  {10.1088/0004-637x/814/2/88}, 814, 88

\bibitem[\protect\citeauthoryear{Salvetti et~al.,}{Salvetti
  et~al.}{2017}]{Salvetti2017}
Salvetti D.,  et~al., 2017, \mn@doi [Monthly Notices of the Royal Astronomical
  Society] {10.1093/mnras/stx1247}, 470, 466

\bibitem[\protect\citeauthoryear{{Saz Parkinson}, Xu, Yu, Salvetti, Marelli  \&
  Falcone}{{Saz Parkinson} et~al.}{2016}]{SazParkinson2016}
{Saz Parkinson} P.~M.,  Xu H.,  Yu P. L.~H.,  Salvetti D.,  Marelli M.,
  Falcone A.~D.,  2016, \mn@doi [The Astrophysical Journal]
  {10.3847/0004-637x/820/1/8}, 820, 8

\bibitem[\protect\citeauthoryear{Strader, Chomiuk, Sonbas, Sokolovsky, Sand,
  Moskvitin  \& Cheung}{Strader et~al.}{2014}]{Strader2014}
Strader J.,  Chomiuk L.,  Sonbas E.,  Sokolovsky K.,  Sand D.~J.,  Moskvitin
  A.~S.,   Cheung C.~C.,  2014, \mn@doi [The Astrophysical Journal]
  {10.1088/2041-8205/788/2/l27}, 788, L27

\bibitem[\protect\citeauthoryear{Strader et~al.,}{Strader
  et~al.}{2019}]{Strader2019}
Strader J.,  et~al., 2019, \mn@doi [The Astrophysical Journal]
  {10.3847/1538-4357/aafbaa}, 872, 42

\bibitem[\protect\citeauthoryear{Süveges}{Süveges}{2014}]{Suveges}
Süveges M.,  2014, \mn@doi [Monthly Notices of the Royal Astronomical Society]
  {10.1093/mnras/stu372}, 440, 2099

\bibitem[\protect\citeauthoryear{Torres, Ji, Li, Papitto, Rea, de
  O{\~{n}}a~Wilhelmi  \& Zhang}{Torres et~al.}{2017}]{Torres2017}
Torres D.~F.,  Ji L.,  Li J.,  Papitto A.,  Rea N.,  de O{\~{n}}a~Wilhelmi E.,
   Zhang S.,  2017, \mn@doi [The Astrophysical Journal]
  {10.3847/1538-4357/836/1/68}, 836, 68

\bibitem[\protect\citeauthoryear{VanderPlas \&
  Ivezic{\textasciiacute}}{VanderPlas \&
  Ivezic{\textasciiacute}}{2015}]{VanderPlas2015}
VanderPlas J.~T.,  Ivezic{\textasciiacute} {\v{Z}}.,  2015, \mn@doi [The
  Astrophysical Journal] {10.1088/0004-637x/812/1/18}, 812, 18

\bibitem[\protect\citeauthoryear{VanderPlas, Connolly, Ivezic  \&
  Gray}{VanderPlas et~al.}{2012}]{VanderPlas2012}
VanderPlas J.~T.,  Connolly A.~J.,  Ivezic Z.,   Gray A.,  2012, in 2012
  Conference on Intelligent Data Understanding. {IEEE},
  \mn@doi{10.1109/cidu.2012.6382200}, \url
  {https://doi.org/10.1109/cidu.2012.6382200}

\bibitem[\protect\citeauthoryear{{Wang} et~al.,}{{Wang}
  et~al.}{2018}]{Wang2018}
{Wang} P.,  et~al., 2018, The Astronomer's Telegram, \href
  {https://ui.adsabs.harvard.edu/abs/2018ATel11584....1W} {11584, 1}

\bibitem[\protect\citeauthoryear{Yao, Manchester  \& Wang}{Yao
  et~al.}{2017}]{Yao2017}
Yao J.~M.,  Manchester R.~N.,   Wang N.,  2017, \mn@doi [The Astrophysical
  Journal] {10.3847/1538-4357/835/1/29}, 835, 29

\bibitem[\protect\citeauthoryear{{Zyuzin}, {Karpova}  \& {Shibanov}}{{Zyuzin}
  et~al.}{2018}]{Zyuzin2018}
{Zyuzin} D.~A.,  {Karpova} A.~V.,   {Shibanov} Y.~A.,  2018, \mn@doi [\mnras]
  {10.1093/mnras/sty359}, \href
  {https://ui.adsabs.harvard.edu/abs/2018MNRAS.476.2177Z} {476, 2177}

\makeatother
\end{thebibliography}

\label{lastpage}
\end{document}